\documentclass[twoside,twocolumn,9pt]{article}
\usepackage{extsizes}
\usepackage[super,sort&compress,comma]{natbib} 
\usepackage[version=3]{mhchem}
\usepackage[left=1.5cm, right=1.5cm, top=1.785cm, bottom=2.0cm]{geometry}
\usepackage{balance}
\usepackage{mathptmx}
\usepackage{sectsty}
\usepackage{graphicx} 
\usepackage{lastpage}
\usepackage[format=plain,justification=justified,singlelinecheck=false,font={stretch=1.125,small,sf},labelfont=bf,labelsep=space]{caption}
\usepackage{float}
\usepackage{fancyhdr}
\usepackage{fnpos}
\usepackage[english]{babel}
\addto{\captionsenglish}{%
  
}
\usepackage{array}
\usepackage{droidsans}
\usepackage{charter}
\usepackage[T1]{fontenc}
\usepackage[usenames,dvipsnames]{xcolor}
\usepackage{setspace}
\usepackage[compact]{titlesec}
\usepackage{hyperref}

\usepackage{amssymb}
\usepackage{amsmath}
\usepackage{bm}
\usepackage{graphicx}
\usepackage{hyperref}
\usepackage{xcolor}
\usepackage{soul}

\usepackage[utf8]{inputenc}

\definecolor{cream}{RGB}{222,217,201}

\begin{document}

\pagestyle{fancy}
\thispagestyle{plain}
\fancypagestyle{plain}{
\renewcommand{\headrulewidth}{0pt}
}

\makeFNbottom
\makeatletter
\renewcommand\LARGE{\@setfontsize\LARGE{15pt}{17}}
\renewcommand\Large{\@setfontsize\Large{12pt}{14}}
\renewcommand\large{\@setfontsize\large{10pt}{12}}
\renewcommand\footnotesize{\@setfontsize\footnotesize{7pt}{10}}
\makeatother

\renewcommand{\thefootnote}{\fnsymbol{footnote}}
\renewcommand\footnoterule{\vspace*{1pt}%
\color{cream}\hrule width 3.5in height 0.4pt \color{black}\vspace*{5pt}} 
\setcounter{secnumdepth}{5}

\makeatletter 
\renewcommand\@biblabel[1]{#1}            
\renewcommand\@makefntext[1]%
{\noindent\makebox[0pt][r]{\@thefnmark\,}#1}
\makeatother 
\renewcommand{\figurename}{\small{Fig.}~}
\sectionfont{\sffamily\Large}
\subsectionfont{\normalsize}
\subsubsectionfont{\bf}
\setstretch{1.125} 
\setlength{\skip\footins}{0.8cm}
\setlength{\footnotesep}{0.25cm}
\setlength{\jot}{10pt}
\titlespacing*{\section}{0pt}{4pt}{4pt}
\titlespacing*{\subsection}{0pt}{15pt}{1pt}

\fancyfoot{}
\fancyfoot[LO,RE]{\vspace{-7.1pt}\includegraphics[height=9pt]{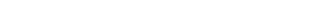}}
\fancyfoot[CO]{\vspace{-7.1pt}\hspace{13.2cm}\includegraphics{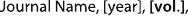}}
\fancyfoot[CE]{\vspace{-7.2pt}\hspace{-14.2cm}\includegraphics{RF.pdf}}
\fancyfoot[RO]{\footnotesize{\sffamily{1--\pageref{LastPage} ~\textbar  \hspace{2pt}\thepage}}}
\fancyfoot[LE]{\footnotesize{\sffamily{\thepage~\textbar\hspace{3.45cm} 1--\pageref{LastPage}}}}
\fancyhead{}
\renewcommand{\headrulewidth}{0pt} 
\renewcommand{\footrulewidth}{0pt}
\setlength{\arrayrulewidth}{1pt}
\setlength{\columnsep}{6.5mm}
\setlength\bibsep{1pt}

\makeatletter 
\newlength{\figrulesep} 
\setlength{\figrulesep}{0.5\textfloatsep} 

\newcommand{\topfigrule}{\vspace*{-1pt}%
\noindent{\color{cream}\rule[-\figrulesep]{\columnwidth}{1.5pt}} }

\newcommand{\botfigrule}{\vspace*{-2pt}%
\noindent{\color{cream}\rule[\figrulesep]{\columnwidth}{1.5pt}} }

\newcommand{\dblfigrule}{\vspace*{-1pt}%
\noindent{\color{cream}\rule[-\figrulesep]{\textwidth}{1.5pt}} }

\makeatother

\twocolumn[
  \begin{@twocolumnfalse}
{\includegraphics[height=30pt]{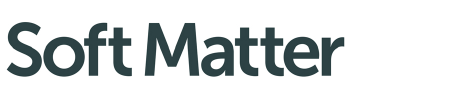}\hfill\raisebox{0pt}[0pt][0pt]{\includegraphics[height=55pt]{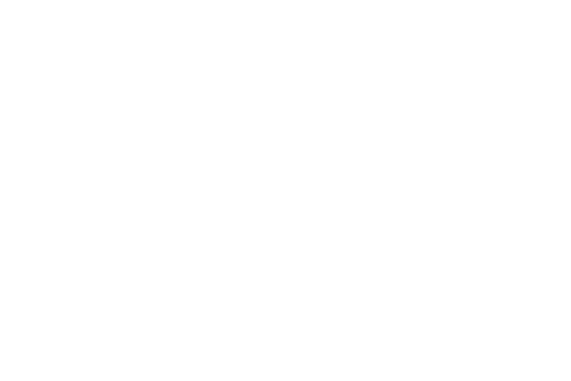}}\\[1ex]
\includegraphics[width=18.5cm]{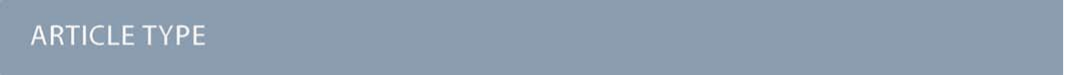}}\par
\vspace{1em}
\sffamily
\begin{tabular}{m{4.5cm} p{13.5cm} }

\includegraphics{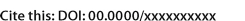} & \noindent\LARGE{\textbf{Analytical model for the motion and interaction of two-dimensional active nematic defects}} \\
\vspace{0.3cm} & \vspace{0.3cm} \\

 & \noindent\large{Cody D. Schimming$^*$, C. J. O. Reichhardt, and C. Reichhardt} \\

\includegraphics{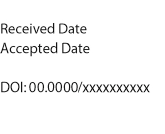} & \noindent\normalsize{We develop an approximate, analytical model for the velocity of defects in active nematics by combining recent results for the velocity of topological defects in nematic liquid crystals with the flow field generated from individual defects in active nematics.
Importantly, our model takes into account the long-range interactions between defects that result from the flows they produce as well as the orientational coupling between defects inherent in nematics.
\textcolor{black}{Our work complements previous studies of active nematic defect motion by introducing a linear approximation that allows us to treat defect interactions as two-body interactions and incorporates the hydrodynamic screening length as a tuning parameter.}
We show that the model can analytically predict bound states between two $+1/2$ winding number defects, effective attraction between two $-1/2$ defects, and the scaling of a critical unbinding length between $\pm 1/2$ defects with activity. 
The model also gives predictions for the trajectories of defects, such as the scattering of $+1/2$ defects by $-1/2$ defects at a critical impact parameter that depends on activity. 
In the presence of circular confinement, the model predicts a braiding motion for three $+1/2$ defects that was recently seen in experiments,
\textcolor{black}{as well as stable and ergodic trajectories for four or more defects.}
} \\

\end{tabular}

 \end{@twocolumnfalse} \vspace{0.6cm}

  ]

\renewcommand*\rmdefault{bch}\normalfont\upshape
\rmfamily
\section*{}
\vspace{-1cm}


\footnotetext{\textit{Theoretical Division and Center for Nonlinear Studies, Los Alamos National Laboratory, Los Alamos, New Mexico, 87545, USA}}
\footnotetext{\textit{E-mail: cschimm2@jh.edu}}




\section{Introduction}

Topological defects are present in many physical systems, including solids, superfluids and superconductors, magnetic materials, complex fluids, soft and biological materials, and the cosmos \cite{chaikin95,kibble97,pismen99,kleman08,saw17,Maroudas-Sacks21}. 
Although the physical properties and scales of all of these systems vary widely, the progenitor of topological defects in each case is broken symmetry. 
Topological defects manifest as singularities in the broken symmetry order parameter, and, as such, take the form of quantized points, lines, surfaces, or hypersurfaces. 
The quantization of topological defects thus allows for an alternative picture of the structure and dynamics of these systems; namely, viewing topological defects as interacting particles. 
For some systems, such as vortices in type-II superconductors or Skyrmions in chiral magnets, the ``particle model'' approach has been successfully used to describe the behavior of the system \cite{Shi-Zeng13,Reichhardt16}. 
In other systems, such as dislocations in solids and disclinations in liquid crystals, only recently have the required theoretical advances
been made that permit defects
to be treated as particles
\textcolor{black}{by relating them to gradients of the continuum order parameter.}
\cite{tang17,audun18,Shankar19,angheluta21,long21,schimming22,skogvoll22,schimming23}.

Here, we focus on topological defects in two-dimensional active nematic liquid crystals. 
Active nematic liquid crystals, or simply ``active nematics,'' are fluids in which anisotropic constituents individually generate forces leading to collective macroscopic flows \cite{marchetti13,doo18}. 
Active nematics are inherently out-of-equilibrium and exhibit chaotic flows on large scales \cite{DeCamp15,giomi15,doo16b,lemma19}.
They also exhibit spontaneous positive-negative defect pair production with self-advecting positive defects \cite{giomi13,shankar18,Shankar19}.

Much of the theoretical and computational work surrounding active nematics has focused on a continuum model for the time dependence of the nematic tensor order parameter $\mathbf{Q}$ and the fluid velocity field $\mathbf{v}$ \cite{marenduzzo07,marchetti13,doo18}. 
There has been recent interest, however, in the dynamics of the topological defects themselves for their role in colloidal assembly, microfluidic cargo transport, morphogenesis, logic operations, and optimal mixing \cite{lazo14,re:conklin17,genkin17,li17,saw17,Tan19,copenhagen21,Maroudas-Sacks21,figueroa22,RZhang22,Serra23,Shankar24,Memarian24,Mitchell24}. 
The behavior of defects is typically inferred from solutions to the continuum equations for the order parameter,
but the nonlinearity of the time evolution equations limits the progress
that can be made in obtaining analytical results,
\textcolor{black}{though recent progress has been made by using low-order expansions of the equations of motion of $\mathbf{Q}$ \cite{Cortese18,ZhangYi20}.}

\textcolor{black}{There have been considerable efforts to understand the flow fields produced by individual defects in active nematics, both with and without substrate friction \cite{giomi13,Pismen13,giomi14,giomi15,Pismen17,angheluta21,ronning22,Ronning23}.}
\textcolor{black}{In many of these studies, defect motion is assumed to be dominated by the advection of the defect by the flows they produce.}
\textcolor{black}{Additionally, because of the reliance on exact solutions for the fluid flows, understanding the motion of multiple interacting defects through them remains a challenge.}
\textcolor{black}{Other recent studies of defect dynamics on curved surfaces have made the simplifying assumptions that positive defects are self-propelled, while negative defects are passive and only interact through Coulomb potentials and surface curvature coupling \cite{Keber14,ellis18}.}
\textcolor{black}{These models introduce noise (an experimental reality) and allow for independent orientations of defects, however, they}
do not include the coupling of defect orientations with relative positions, or the advection and shear flow interactions inherent in active nematics \cite{Pearce21b,Head24}.
\textcolor{black}{Recent work has also explored the effect of orientational noise on the unbinding of defects\cite{shankar18} and developed a hydrodynamic description of many interacting defects \cite{Shankar19,angheluta21,Shankar24}.}
\textcolor{black}{The hydrodynamic descriptions are suitable for systems with large numbers of active defects in the turbulent regime, but do not give descriptions of the trajectories of individual defects.}

In this work, we combine recent analytical results for the flow field around a defect in an active nematic with an analytical framework to calculate the velocity of defects given the time dependence of the order parameter \cite{angheluta21,ronning22,schimming23}. 
Using these elements, we posit an analytical model for the velocity of defects in an active nematic with many defects. 
\textcolor{black}{A critical component to our model is a linear perturbation approximation for the $\mathbf{Q}$-tensor order parameter near defect cores.}
\textcolor{black}{Using this,} we show that 
the defect velocity \textcolor{black}{results from two-body interactions with other defects, and} may be decomposed into three parts: the Coulomb interaction between defects, advection from fluid flows generated by other defects, and deflection by shear flows generated by other defects. 
We analytically probe the interaction between two defects and show that the model is able to predict 
\textcolor{black}{phenomena observed in continuum simulations and experiments, and predicted by other analytical models such as}
bound states between positive defects and deflections of positive defects by negative defects  \cite{angheluta21,Vafa22}.
\textcolor{black}{We also show that the model predicts an orientation dependent, effective attraction between negative defects and that the critical unbinding length scales proportional to the active length, both of which have not been shown in previous analytical models to our knowledge.}
We study the predicted trajectories for multiple defects in a circularly confined system and identify several potential behaviors, such as stable configurations, periodic dynamics, ergodic trajectories, and defect braiding similar to that observed in recent experiments \cite{Memarian24}. 
We also compare the analytical model to a continuum model for the case of defect coarsening in a confined system.
Finally, we summarize our results and discuss potential future extensions of the model and challenges that remain. 
\textcolor{black}{Our work complements the previous efforts to understand active nematic defect motion by explicitly including the hydrodynamic screening length as a tuning parameter, and allowing for the study of multiple interacting defects through a key approximation.}

\section{Velocity of Active Nematic Defects} \label{sec:DefectVelocity}

In this section, we derive an analytical expression for the velocity of defects in an active nematic in the presence of other defects. 
We present the primary equations in the main text while relegating some of the details to the Appendix.

\subsection{Defect velocity in an arbitrary flow field}

The local orientational order in a nematic phase may be characterized by a unit vector field called the director, $\mathbf{\hat{n}}(\mathbf{r})$. 
Nematic phases have apolar symmetry, so $-\mathbf{\hat{n}} \equiv \mathbf{\hat{n}}$. 
In two-dimensional nematics, topological defects are points of orientational singularity, and, hence, are points where $\mathbf{\hat{n}}$ cannot be defined. 
To identify disclinations, the winding number or ``topological charge'' is computed:
\begin{equation}
    m = \frac{1}{2 \pi}\oint_C d\phi
\end{equation}
where $C$ is a closed loop in the nematic and $\phi$ is the angle of $\mathbf{n}$ with respect to an arbitrary reference axis. 
Since $\mathbf{\hat{n}}$ and $-\mathbf{\hat{n}}$ represent the same state, it is possible for defects of charge $m = n/2$, where $n$ is an integer, to exist. 
The lowest energy nontrivial defects have charge $m = \pm 1/2$ \cite{deGennes75}. 
Here, we will only consider these half integer defects.

For systems with many defects, it is common to represent the nematic configuration using a rank-two tensor, $\mathbf{Q}$. 
In two dimensions, $\mathbf{Q}$ may be parameterized as $\mathbf{Q} = S\left[\mathbf{\hat{n}} \otimes \mathbf{\hat{n}} - (1/2)\mathbf{I}\right]$ where $S$ is a scalar representing the degree of local order. 
The tensor order parameter characterization has the benefit that topological defects are no longer singularities and are instead regularized so that $S$ varies continuously and $S \to 0$ at the defect cores. 
The apolar symmetry is also inherently present in $\mathbf{Q}$ due to its quadratic dependence on $\mathbf{\hat{n}}$.

Recently, an exact expression for the velocity of defects in terms of derivatives of $\mathbf{Q}$ was derived in both two and three dimensions \cite{angheluta21,schimming23}.
\textcolor{black}{The derivation, of which we give more details in the Appendix, is based on the conservation of topological charge for systems with a continuous, broken symmetry order parameter, first explored by Halperin\cite{halperin81} and then later developed by Mazenko\cite{liu92,mazenko97,mazenko99}.}
In two dimensions it reads
\begin{equation} \label{eqn:DefectVelocity}
    v_i = \left. \mp 4 \frac{\varepsilon_{i j} \varepsilon_{\mu \nu} \partial_t Q_{\mu \alpha} \partial_j Q_{\nu \alpha}}{\varepsilon_{k \ell}\varepsilon_{\eta \tau}\partial_k Q_{\eta \beta} \partial_{\ell} Q_{\tau \beta}} \right|_{\mathbf{r} = \mathbf{r}_0}
\end{equation}
where $\bm{\varepsilon}$ is the two-dimensional antisymmetric matrix, $\partial_t \equiv \partial/\partial t$, $\partial_j \equiv \partial/\partial x_j$, and summation on repeated indices is assumed. 
Importantly, the velocity of the defect only depends on derivatives of $\mathbf{Q}$ at the location of the defect core, $\mathbf{r} = \mathbf{r}_0$. 
While Eq.~\eqref{eqn:DefectVelocity} is exact, $\mathbf{Q}$ may be complicated and non-linear, leading to difficulties in obtaining analytical results for defect velocities. 
Further, the time dependence of $\mathbf{Q}$ must also be known
in order to use Eq.~\eqref{eqn:DefectVelocity}.

To obtain an approximate analytical expression for the velocity of defects, we assume the Beris-Edwards model for the time evolution of $\mathbf{Q}$ in the presence of a fluid flow $\mathbf{u}$ \cite{beris94}:
\begin{equation} \label{eqn:QEvo}
    \partial_t Q_{\mu \alpha} = -\left(\mathbf{u} \cdot \nabla\right) Q_{\mu \alpha} + S_{\mu \alpha} - \frac{1}{\gamma} \frac{\delta F}{\delta Q_{\mu \alpha}}
\end{equation}
where $\gamma$ is a rotational viscosity and
\begin{multline}
    \mathbf{S} = \left(\lambda \mathbf{E} + \bm{\Omega}\right)\left(\mathbf{Q} + \frac{1}{2}\mathbf{I}\right) + \left(\mathbf{Q} + \frac{1}{2}\mathbf{I}\right)\left( \lambda\mathbf{E} - \bm{\Omega}\right) \\
    - 2\lambda\left(\mathbf{Q} + \frac{1}{2}\right)\left(\nabla \mathbf{u} : \mathbf{Q}\right)
\end{multline}
is a generalized tensor advection. 
Here $\mathbf{E}$ is the strain rate tensor, $\bm{\Omega}$ is the vorticity tensor and $\lambda$ is a dimensionless parameter that characterizes the tendency of nematogens to either tumble or align due to shear flow \cite{leslie66}. 
The free energy, $F$, is chosen so that its bulk density is analytic in $\mathbf{Q}$ (e.g. Landau-de Gennes \cite{deGennes75}) and has a one-constant elastic density $f_E = L |\nabla \mathbf{Q}|^2$. For the remainder of the paper we work in dimensionless units by scaling lengths by the nematic correlation length $\xi = \sqrt{k L/C}$ and times by the nematic relaxation time $\tau = \gamma/(C\xi^2)$, where $k$ is a dimensionless parameter that determines the size of defects and relative strength of elastic forces and $C$ is a characteristic energy density scale of the free energy.

Equation \eqref{eqn:QEvo} may be substituted into Eq. \eqref{eqn:DefectVelocity} to yield an equation for the defect velocity in terms of spatial derivatives of $\mathbf{Q}$ only. 
To approximate spatial derivatives at the location of the defect core, we use a linear core approximation for a nematic defect in the presence of other defects \cite{long21,schimming23}. 
In the absence of flow, $\mathbf{u} = 0$, the velocity of the $j$th defect reduces to
\begin{equation} \label{eqn:CoulombVelocity}
    \mathbf{v}_j^{\text{Coulomb}} = \frac{16}{k}\sum_{i\neq j} \frac{m_i m_j \mathbf{r}_{ij}}{|\mathbf{r}_{ij}|^2}
\end{equation}
where $\mathbf{r}_{ij} = \mathbf{r}_j- \mathbf{r}_i$ is the vector pointing from defect $i$ to defect $j$. 
Equation \eqref{eqn:CoulombVelocity} is precisely the Coulomb interaction between defects in passive nematic liquid crystals \cite{deGennes75}. 
In what follows, we will set $k = 4$ unless otherwise specified. 
In the presence of a flow field, $\mathbf{u} \neq 0$, Eq. \eqref{eqn:DefectVelocity} gives two additional contributions to the defect velocity. The first is advection by the flow:
\begin{equation} \label{eqn:AdvectionVelocity}
    \mathbf{v}_j^{\text{Advection}} = \mathbf{u}(\mathbf{r} = \mathbf{r}_j).
\end{equation}
Equation \eqref{eqn:AdvectionVelocity} shows that defects are simply advected by the flow velocity at the location of the defect core, which is an expected and intuitive result.
The second contribution comes from the deflection of defects in shear flow and has a more complicated form. 
To simplify the notation, we introduce the unit vector 
\begin{align}
    \mathbf{\hat{p}}_j &= \left(\cos2\varphi_j,\,\sin2\varphi_j\right) \label{eqn:phat} \\
    \varphi_j &= \lim_{\epsilon \to 0} \phi(x_j + \epsilon,y_j) \label{eqn:angle}
\end{align}
where $\phi(x,y)$ is the angle of the director, $\mathbf{\hat{n}}(x,y)$, with respect to the $x$-axis and the index $j$ refers to the $j$th defect. 
Thus $\varphi_j$ is the angle of the director field just outside the defect core of the $j$th defect in the $+x$ direction and $\mathbf{\hat{p}}_j$ gives information about the ``orientation'' of the $j$th defect. 
We also write the shear rate tensor as
\begin{equation}
    \begin{pmatrix}
        E_1 & E_2 \\
        E_2 & -E_1
    \end{pmatrix}
\end{equation}
which we assume to be traceless since we will only consider incompressible flows. 
Then
the last contribution to defect velocity is
\begin{multline} \label{eqn:ShearVelocity}
    \mathbf{v}_j^{\text{Shear}} = -\lambda\left[\left(E_1 \cos2\varphi_j + E_2 \sin2\varphi_j\right)\mathbf{\hat{x}} \right. \\ \left. + 2m_j\left(-E_1 \sin2\varphi_j + E_2\cos2\varphi_j\right)\mathbf{\hat{y}}\right]
\end{multline}
where $E_1$ and $E_2$ are computed at $\mathbf{r} = \mathbf{r}_j$. 
We note that Eq. \eqref{eqn:ShearVelocity} depends on the charge of the defect and, in particular, the $y$-component of the velocity changes sign for different signed charges. 
For the remainder of the paper we will set $\lambda = 1$.

\begin{figure}
\centering
    \includegraphics[width = \columnwidth]{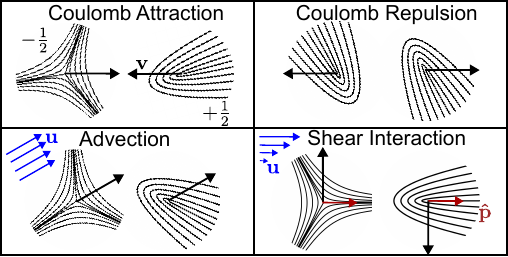}
\caption{Illustration of defect motion in two-dimensional nematics. Coulomb forces result from elastic interactions while advection and shear interactions only occur when a flow (blue arrows) is present.}
    \label{fig:DefectVelocities}
\end{figure}

The defect velocity in the presence of flow is the sum of Eqs. \eqref{eqn:CoulombVelocity}, \eqref{eqn:AdvectionVelocity}, and \eqref{eqn:ShearVelocity}. 
In Fig. \ref{fig:DefectVelocities} we show an illustration of each contribution. 
In the next section we use recent analytical calculations of the flow field created by a defect in an active nematic to give an analytical expression for the advection and shear contributions to defect velocity.

\subsection{Flow from active nematic defects}

In this section we review the analytical results from Ref. \cite{ronning22} for the flow field around a defect in an active nematic and combine them with the results of the previous section to obtain an expression for the velocity of a defect in an active nematic. 
We consider solutions of the incompressible Stokes equation with active nematic stress:
\begin{align}
    \eta \nabla^2 \mathbf{u} - \Gamma \mathbf{u} &= \nabla p + \tilde{\alpha} \nabla \cdot \mathbf{Q} \label{eqn:Stokes} \\
    \nabla \cdot \mathbf{u} &= 0 \nonumber
\end{align}
where $\eta$ is the fluid viscosity, $\Gamma$ is the friction coefficient with the substrate, $p$ is the fluid pressure, and $\tilde{\alpha}$ is the strength of the active force.
Dividing Eq.~\eqref{eqn:Stokes} by $\Gamma$ yields the parameters $\alpha \equiv \tilde{\alpha} / \Gamma$, which characterizes the magnitude of the active force and hence determines the magnitude of the flows, and $\ell_h \equiv \sqrt{\eta/\Gamma}$, which is the hydrodynamic screening length that characterizes the spatial extent of the flows. 
\textcolor{black}{We note that (as is done in Ref. \cite{ronning22}) we have neglected the effect of elastic stresses in Eq.~\eqref{eqn:Stokes}.
To include these terms in our model, we would require an analytical solution to Eq.~\eqref{eqn:Stokes} that includes them, which, to our knowledge, has not been found.
Nevertheless, it has been shown that the active stress dominates elastic stresses in other solutions of the Stokes equation for nematic defects \cite{giomi13,Pismen13,giomi14}.
In a passive nematic, it would likely be much more important to include these terms, as the elastic backflow has been shown to alter the interaction of defects when active stresses are not present \cite{toth02}.}

We will assume that the total flow at the core of a defect is the sum of all flow fields created by defects. 
We justify this assumption by first noting that the director field in a system of many defects is determined by the Euler-Lagrange equation $\nabla^2 \phi = 0$. 
This equation is linear, so the full solution is the sum of each individual defect solution. In particular, we will assume the director structure is given by the minimal energy solution, so that there are no torques on defects due to elastic forces \cite{vromans16,tang17}. 
Physically, this corresponds to assuming that the nematic relaxation time is much smaller than the active force time scale. 
The flow field created by a defect is sourced by the term $\alpha \nabla \cdot \mathbf{Q}$ in the rescaled Eq.~\eqref{eqn:Stokes}. 
Away from the core of other defects, this term depends only on the director, and, under our linear core assumptions, is given by the sum of all other defects. 
We thus approximate the total active force by the sum of active forces produced by each defect. 
This approximation is better when defects are well separated compared to their core size, and will begin to break down as defect cores overlap. 
Nevertheless, these approximations and assumptions allow us to make analytical progress in treating interactions between multiple active nematic defects,
\textcolor{black}{since they allow us to treat the velocities as arising from two-body interactions.}

\begin{figure}
\centering
    \includegraphics[width = \columnwidth]{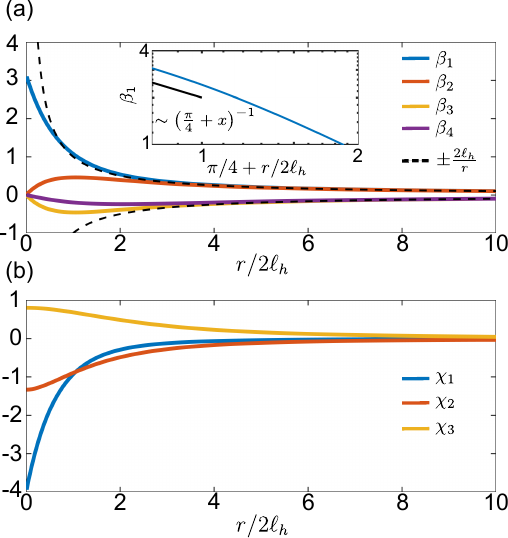}
\caption{(a) Radial parts $\beta_1$, $\beta_2$, $\beta_3$, and $\beta_4$
of the flow generated by active nematic defects plotted vs
$r/2\ell_h$.
\textcolor{black}{Inset: Log-log plot of $\beta_1$ vs $\pi/4 + r/2\ell_h$.}
(b) Radial parts $\chi_1$, $\chi_2$, and $\chi_3$ of the shear rate generated by active nematic defects plotted vs $r/2\ell_h$.}
    \label{fig:VelocityFunctions}
\end{figure}

In active nematics, $+1/2$ defects are much more motile than $-1/2$ defects. 
This is because they generate a nonzero flow at their core \cite{ronning22}
\begin{equation}
    \mathbf{u}^+(0) = -\frac{\pi \alpha}{4 \ell_h}\mathbf{\hat{p}}.
\end{equation}
On the other hand, $-1/2$ defects generate zero flow at their core, $\mathbf{u}^-(0) = 0$, though they still move due to interactions with other defects and flows. 
The flow field away from the core of $\pm 1/2$ defects is given by \cite{ronning22}
\begin{multline}
    u_x^+(r,\theta,\varphi) = -\frac{\alpha}{4\ell_h}\left[\cos(2\varphi)\beta_1\left(\frac{r}{2\ell_h}\right) \right. \\ \left. + \cos(2\theta - 2\varphi)\beta_2\left(\frac{r}{2\ell_h}\right)\right]
\end{multline}
\begin{multline}
    u_y^+(r,\theta,\varphi) = -\frac{\alpha}{4\ell_h}\left[\sin(2\varphi)\beta_1\left(\frac{r}{2\ell_h}\right) \right. \\ \left. + \sin(2\theta - 2\varphi)\beta_2\left(\frac{r}{2\ell_h}\right)\right]
\end{multline}
\begin{multline}
    u_x^-(r,\theta,\varphi) = -\frac{\alpha}{4\ell_h}\left[\cos(2\theta - 2\varphi)\beta_3\left(\frac{r}{2\ell_h}\right) \right. \\  \left. + \cos(4\theta - 2\varphi)\beta_4\left(\frac{r}{2\ell_h}\right)\right]
\end{multline}
\begin{multline}
    u_y^-(r,\theta,\varphi) = -\frac{\alpha}{4\ell_h}\left[-\sin(2\theta - 2\varphi)\beta_3\left(\frac{r}{2\ell_h}\right) \right. \\ \left. + \sin(4\theta - 2\varphi)\beta_4\left(\frac{r}{2\ell_h}\right)\right]
\end{multline}
where $r$ and $\theta$ are polar coordinates and $\varphi$ gives the orientation of the defect, given by Eq. \eqref{eqn:angle}. 
The functions $\beta_i$ are plotted in Fig. \ref{fig:VelocityFunctions}(a) and the functional forms are given in the Appendix. 
As shown in Fig. \ref{fig:VelocityFunctions}(a), at large arguments $\beta_i(x) \sim 1/x$, while at smaller arguments the behavior is more complicated and even nonmonotonic in the case of $\beta_{2-4}$.

$E_1$ and $E_2$ may be computed from gradients of the velocity. 
For $\pm 1/2$ defects these are given by
\begin{multline}
    E_1^+(r,\theta,\varphi) = -\frac{\alpha}{16 \ell_h^2}\left[\cos(\theta + 2\varphi)\chi_1\left(\frac{r}{2\ell_h}\right) \right. \\ \left. + \cos(3\theta - 2\varphi)\chi_2\left(\frac{r}{2\ell_h}\right)\right]
\end{multline}
\begin{multline}
    E_2^+(r,\theta,\varphi) = -\frac{\alpha}{16 \ell_h^2}\left[\sin(\theta + 2\varphi)\chi_1\left(\frac{r}{2\ell_h}\right) \right. \\ \left. + \sin(3\theta - 2\varphi)\chi_2\left(\frac{r}{2\ell_h}\right)\right]
\end{multline}
\begin{multline}
    E_1^-(r,\theta,\varphi) = -\frac{\alpha}{16 \ell_h^2}\left[\cos(\theta - 2\varphi)\chi_1\left(\frac{r}{2 \ell_h}\right) \right. \\ \left. + \cos(5\theta - 2\varphi)\chi_3\left(\frac{r}{2\ell_h}\right)\right]
\end{multline}
\begin{multline}
    E_2^-(r,\theta,\varphi) = -\frac{\alpha}{16 \ell_h^2}\left[-\sin(\theta - 2\varphi)\chi_1\left(\frac{r}{2\ell_h}\right) \right. \\ \left. + \sin(5\theta - 2\varphi)\chi_3\left(\frac{r}{2 \ell_h}\right)\right]
\end{multline}
where
\begin{align*}
    \chi_1(x) &= \frac{d\beta_1(x)}{dx} = \frac{d\beta_3(x)}{dx} + \frac{2}{x}\beta_3(x) \\
    \chi_2(x) &= \frac{d\beta_2(x)}{dx} - \frac{2}{x}\beta_2(x) \\
    \chi_3(x) &= \frac{d\beta_4(x)}{dx} - \frac{4}{x}\beta_4(x)
\end{align*}
are plotted in Fig. \ref{fig:VelocityFunctions}(b). 
For large arguments, $\chi_i(x) \sim 1/x^2$, and hence the shear flow interaction will be subdominant when distances are much larger than the hydrodynamic screening length. 
This is the intuitive result for ``dry'' active nematics where $\ell_h \to 0$. 
In this case, the self-propulsion of $+1/2$ defects dominates other terms in the velocity. 

\textcolor{black}{We note that the limit $\ell_h \to \infty$ is not a sensible limit to take in the model.
This is because, in this limit, the substrate friction goes to zero and solutions for the two-dimensional Stokes equation in an infinite domain diverge.
This does not contradict experiments because substrate friction always plays a role, or in continuum simulations that set $\Gamma = 0$ since the finite size of domains always acts as an effective screening length on the system.
A more sensible limit to examine in our model is how the flow solutions behave in the limit $r << \ell_h$.
While the form of $\beta_i$ and $\chi_i$ is very complicated in this limit, we use the first two terms of $\beta_1$ (given in the Appendix) to estimate that $\beta_1(x) \sim (\pi^2/4)/(\pi/4 + x)$ for small $x$. 
Indeed, for small $r/2\ell_h$ we find this to be a satisfactory approximation, as we show in the inset of Fig.~\ref{fig:VelocityFunctions}(a).}

The velocity of an individual defect is found by treating all other defects as point sources for the flow at the location of the defect of interest. 
Due to flows throughout the nematic, the actual nematic configuration may be more complicated; however, we do not consider this here and assume that all flows in the system are due to topological defects. 
In the limit of small relaxation time compared to the timescale for active forces, this is a good approximation, since any non-topologically protected deformation will quickly be relaxed.

\section{Interaction between two active nematic defects} \label{sec:TwoDefects}

Here we analyze the interaction between active nematic defects with the simplifying assumption that only two defects exist in an infinite domain. 
In each case we set our coordinates so that the defect of interest is at polar coordinates $(r,\theta)$ and the other defect is at $(0,0)$. 
We will also assume ``optimal'' orientation between defects \cite{tang17}. 
This set of orientations minimizes the elastic free energy of the configuration. 
Further, given a global phase $\varphi_0$, the orientation of each defect may be determined by the defect positions:
\begin{multline}
    \varphi_{j_+} = \frac{1}{2}\sum_{i_+ \neq j_+} \arctan\left(\frac{y_{j_+} - y_{i_+}}{x_{j_+} - x_{i_+}}\right) \\ - \frac{1}{2}\sum_{i_-} \arctan\left(\frac{y_{j_+} - y_{i_-}}{x_{j_+} - x_{i_-}}\right) + \varphi_0
\end{multline}
where $i_+$ indexes positive defects, $i_-$ indexes negative defects, and $\varphi_{j_-}$ may be found similarly by summing over negative defects $i_- \neq j_-$.

\subsection{Two $+1/2$ Defects}

Given the approximations and assumptions of the previous section, the velocity of a $+1/2$ defect in the presence of another $+1/2$ defect is 
\begin{multline} \label{eqn:PlusPlusVelocity}
    \mathbf{v}^{(++)} = \left[\frac{1}{r} - \frac{\alpha}{16 \ell_h^2}\chi_1\left(\frac{r}{2\ell_h}\right)\right]\mathbf{\hat{r}} + \frac{\alpha}{4\ell_h}\left[-\pi + \beta_1\left(\frac{r}{2\ell_h}\right)\right]\mathbf{\hat{p}} \\ + \frac{\alpha}{4\ell_h}\beta_2\left(\frac{r}{2\ell_h}\right)\mathbf{\hat{a}} - \frac{\alpha}{16 \ell_h^2}\chi_2\left(\frac{r}{2\ell_h}\right)\mathbf{\hat{b}}
\end{multline}
where $\mathbf{\hat{r}}$ is the unit vector pointing to the defect of interest from the other defect, \textcolor{black}{$\mathbf{\hat{a}} = \left[\cos(\theta - 2\varphi_0),\sin(\theta - 2\varphi_0)\right]$}, and \textcolor{black}{$\mathbf{\hat{b}} = \left[\cos(\theta - 4\varphi_0),\sin(\theta - 4\varphi_0)\right]$}. 
Components of Eq.~\eqref{eqn:PlusPlusVelocity} proportional to $\beta_i$ result from advection while components proportional to $\chi_i$ result from the interaction with shear flows. 
We note that the other defect will have exactly the opposite velocity of the defect under consideration.

Equation \eqref{eqn:PlusPlusVelocity} gives several insights into the motion of two $+1/2$ active defects. 
We first note that the self-propulsion velocity is reduced compared to an
isolated defect, particularly at small distances relative to the screening length (note that $\beta_1 \to \pi$ as $r \to 0$). 
The motion is complicated by terms proportional to vectors $\mathbf{\hat{a}}$ and $\mathbf{\hat{b}}$ which depend on the relative orientation of the defects. 
However, we may analyze this coupling by decomposing the velocity into terms parallel and transverse to the direction between defects:
\begin{multline} \label{eqn:PlusPlusPar}
    \textcolor{black}{v_{||}^+ = \frac{1}{r} - \frac{\alpha}{16\ell_h^2} \chi_1} \\ 
    \textcolor{black}{+ \frac{\alpha}{4\ell_h}\left(-\pi + \beta_1 + \beta_2\right)\cos(2\varphi_0) - \frac{\alpha}{16\ell_h^2}\chi_2 \cos(4\varphi_0)}
\end{multline}
\begin{equation} \label{eqn:PlusPlusPerp}
   \textcolor{black}{ v_{\perp}^+ = \frac{\alpha}{4\ell_h}\left(-\pi + \beta_1 - \beta_2\right)\sin(2\varphi_0) + \frac{\alpha}{16\ell_h^2}\chi_2\sin(4\varphi_0)}
\end{equation}
where we have suppressed the arguments of $\beta_1,\beta_2,\chi_1,\chi_2$ for brevity. 
\textcolor{black}{Equations \eqref{eqn:PlusPlusPar} and \eqref{eqn:PlusPlusPerp} are independent of $\theta$, the angle between defects, and the angular dependence is given by the fixed global phase $\varphi_0$.}
\textcolor{black}{Thus, since $v_{\perp}$ is proportional to the change of angle between defects, the model predicts that defects will rotate around one another as long as $v_\perp \neq 0$.}
\textcolor{black}{We note that the independence of $\theta$ in the defect dynamics is a consequence of the unique rotational symmetry of the far-field director for total topological charge $+1$.}
\textcolor{black}{As we will show in later sections, other defect interactions do not share this property.}

\begin{figure}
\centering
    \includegraphics[width = \columnwidth]{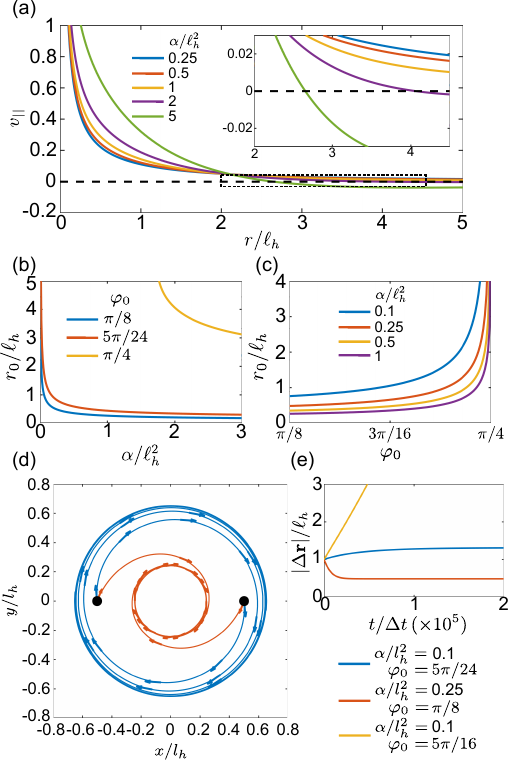}
\caption{(a) Velocity component of a $+1/2$ defect in the presence of another $+1/2$ defect parallel to the direction between defects, $v_{||}^+$ [Eq.~\eqref{eqn:PlusPlusPar}], versus scaled distance between defects $r/\ell_h$. Colors represent different values of $\alpha/\ell_h^2$ with $\ell_h = 10$, $\theta = 0$, $\varphi_0 = \pi/4$, and $\lambda = 1$. The dashed line shows $v_{||}^+ = 0$; $r_0$ is defined as the point at which the velocity crosses this line. (b) Scaled critical distance $r_0/\ell_h$ versus $\alpha/\ell^2_h$ for various values of $\varphi_0$. (c) $r_0 /\ell_h$ versus $\varphi_0$ for various values of $\alpha / \ell_h^2$. (d) Example trajectories for two $+1/2$ defects with initial positions at $x = -0.5\ell_h$ and $x = 0.5 \ell_h$ for $\alpha/\ell_h^2 = 0.1$, $\varphi_0 = 5\pi/24$ (blue) and $\alpha/\ell_h^2 = 0.25$, $\varphi_0 = \pi/8$ (orange). (e) Distance between two $+1/2$ defects versus time step $t / \Delta t$ for the parameters in (d) as well as at $\alpha/\ell_h^2 = 0.1$, $\varphi_0 = 5\pi /16$ (yellow).}
    \label{fig:PlusPlusXVelocity}
\end{figure}

Due to the complicated forms of the flow velocity, it is not immediately obvious whether defects will tend to repel (as they do in passive nematics) or attract. 
Equation \eqref{eqn:PlusPlusPar} predicts that both behaviors are possible, with $\varphi_0 = \pi/4$ acting as a separatrix for the two behaviors. 
In Fig.~\ref{fig:PlusPlusXVelocity}(a) we plot $v_{||}$ as a function of $r/\ell_h$ for
$\varphi_0 = \pi/4$ and $\ell_h = 10$ at 
a variety of activity strengths.
For $\alpha/\ell_h^2 > 1.74$ there is a distance $r_0$ such that $v_{||} = 0$, indicating that defects will tend to rotate around each other at this distance indefinitely. 
In Figs.~\ref{fig:PlusPlusXVelocity}(b,c) we plot the critical distances as a function of $\alpha/\ell_h^2$ (b) and $\varphi_0$ (c). 
For $\varphi_0 < \pi/4$ and $\alpha/\ell_h^2 \lessapprox 1$, $r_0$ diverges at zero activity (as expected), but decreases rapidly with increasing activity. 
For $\alpha/\ell_h^2 > 1$, the critical distance remains roughly constant and small compared to $\ell_h$. 
As a function of $\varphi_0$, $r_0$ remains roughly constant until $\varphi_0 \approx 7\pi/32$, and then increases rapidly as $\varphi_0$ approaches $\pi/4$.

The behavior of $r_0$ may be attributed primarily to the self-advection part of Eq.~\eqref{eqn:PlusPlusPar}, which is 
the dominant term in the equation. It does not depend on the distance between defects, and it contributes most strongly at high activities and smaller $\varphi_0$. 
However, we note that the shear flow interaction contributes most strongly to the behavior near $\varphi_0 = \pi/4$ where the advection terms do not contribute to Eq.~\eqref{eqn:PlusPlusPar}. 
If $\lambda = 0$, $r_0 \to \infty$ for all values of $\alpha$ at $\varphi_0 = \pi/4$.

In Fig.~\ref{fig:PlusPlusXVelocity}(d) we also plot trajectories of defects obtained from numerically integrating Eq.~\eqref{eqn:PlusPlusVelocity} using a forward Euler method with timestep $\Delta t = 0.001$ and $\ell_h = 10$. 
In both computations the defects are initially placed at $(-0.5 \ell_h,0)$ and $(0.5 \ell_h,0)$. 
For one trajectory, we set $\alpha/\ell_h^2 = 0.1$ and $\varphi_0 = 5\pi/24$. 
In this case, $r_0$ is larger than the initial separation, and the defects initially move away from each other to the stable distance. 
For the other trajectory, we set $\alpha/\ell_h^2 = 0.25$ and $\varphi_0 = \pi/8$, corresponding to a critical distance that is smaller than the initial
defect separation.
Here the defects move toward each other in order to reach the stable distance.
In Fig.~\ref{fig:PlusPlusXVelocity}(e) we plot the distance between defects for the computed trajectories illustrated in Fig.~\ref{fig:PlusPlusXVelocity}(d), as well as for trajectories computed for $\alpha/\ell_h^2 = 0.1$ and $\varphi_0 = 5\pi/16 > \pi/4$. 
In this case, while the defects still rotate around one another, the distance between them increases without bound. 

\begin{figure}
\centering
    \includegraphics[width = \columnwidth]{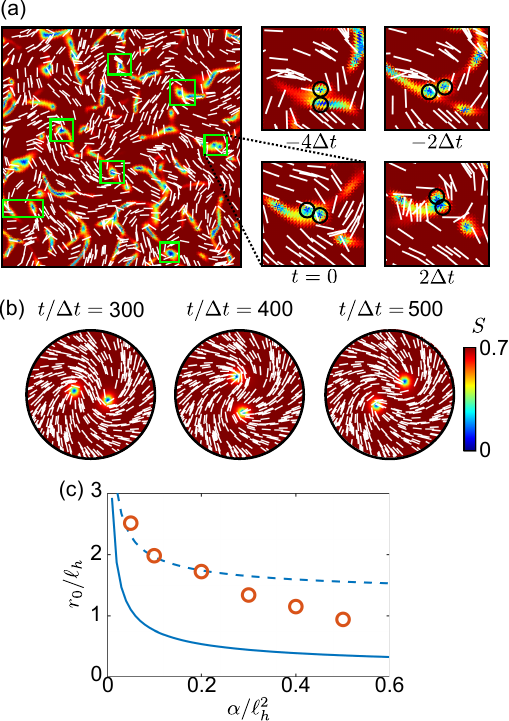}
\caption{(a) Time snapshot of an active nematic continuum simulation with periodic boundary conditions, $\alpha / \ell_h^2 = 2$, and $\ell_h = 10$. 
The color represents the strength of the scalar order parameter $S$ and the white lines indicate the local director. 
The green boxes highlight bound states of two $+1/2$ defects rotating around one another. 
The sub panels show four time snapshots of one of the bound states at $t = - 4 \Delta t$, $t = -2 \Delta t$, $t = 0$, and $t = 2\Delta t$, where $t = 0$ is the time in the main snapshot. \textcolor{black}{The circles indicate the $+1/2$ defects that rotate around one another.} 
(b) Three time snapshots of a confined continuum simulation with only two $+1/2$ defects with $\alpha/\ell_h^2 = 0.2$, $\ell_h = 10$, $\varphi_0 = \pi/8$, and domain radius $R = 30$. 
(c) Scaled stable distance between defects $r_0/\ell_h$ in a continuum simulation for various values of $\alpha/\ell_h^2$ with $\ell_h = 10$ and $\varphi_0 = \pi/8$ (orange points). The dashed line is a reproduction of the analytically derived curve 
\textcolor{black}{from Eq.~\eqref{eqn:PlusPlusPar} (solid line), with a vertical shift of $1.2$.}
}
    \label{fig:PlusPlusContinuum}
\end{figure}

The effective attraction between positive defects caused by hydrodynamic effects can lead to dynamical states in which $+1/2$ defects continuously rotate around one another, or even merge to form a stable higher order $+1$ defect, similar to behaviors predicted with a different analytical model in Ref. \cite{Vafa22}. 
While our calculations above predict this attraction for the simplest case of a system with only two defects, this behavior should, in principle, occur even when other defects are present in the system if the two defects are far from the other defects. 
Indeed, bound states of two $+1/2$ defects may be observed in continuum simulations at high activities, as illustrated in Fig. \ref{fig:PlusPlusContinuum}(a) where we show a time snapshot of the scalar order parameter and director field in a system with periodic boundary conditions, $\alpha/\ell_h^2 = 2$, and $\ell_h = 10$ (for details of the continuum simulations, see the Appendix). 
We have marked the bound states of $+1/2$ defects in green boxes in Fig. \ref{fig:PlusPlusContinuum}(a), and for one of these boxes the subpanels
provide several time snapshots illustrating two bound $+1/2$ defects rotating around one another and eventually merging. 
We note that the bound states are short-lived in the continuum simulation due to interactions with other defects. 
However, it has been shown in numerical simulations that such bound states may be longer lived and ordered if subjected to confinement \cite{norton18,schimming23b,schimming24}. 
Partially stable bound $+1/2$ defect pairs
have also been observed experimentally under circular confinement \cite{opathalage19} and in recently discovered acoustically actuated active nematics with no confinement, where the activity may be much larger than in biologically based active nematics \cite{sokolov24}. 

By simulating a continuum system under circular confinement, we can perform a rough comparison to the scaling of $r_0$ with $\alpha$ shown in Fig.~\ref{fig:PlusPlusXVelocity}(b). 
In Fig.~\ref{fig:PlusPlusContinuum}(b), we plot several time snapshots of a continuum simulation with $\alpha / \ell_h^2 = 0.2$, $\ell_h = 10$, $\varphi_0 = \pi/8$, and a confining radius of $R = 30$. 
In the continuum simulations, two defects rotate around one another at a fixed distance. 
As predicted by the analytical equations, this distance depends on the activity. 
In Fig.~\ref{fig:PlusPlusContinuum}(c) we plot the critical distance $r_0$ versus $\alpha/\ell_h^2$ for several continuum simulations. 
We also plot the analytically derived curve from Fig.~\ref{fig:PlusPlusXVelocity}(b), $r_0(\varphi_0 = \pi/8)/\ell_h + 1.2$, where the vertical shift in the curve is made to highlight the similarities and differences in the shape of the relationship.
Both the continuum simulations and analytic model show a relatively rapid
drop in $r_0$ for small $\alpha$; however, at larger activities $r_0$ falls
off more rapidly with $\alpha$ in the continuum simulations than in the
analytic model,
which we attribute to nonlinear effects produced by
the close proximity of the defect pair
in the continuum simulations. 
When the defects become close enough
together that their cores begin to overlap, our approximations break down. 
In particular, in the continuum model, defects are able to merge to form higher order defects, as is visible in several of the boxed regions in Fig.~\ref{fig:PlusPlusContinuum}(a). 
We surmise that the quantitative difference (the vertical shift between the analytically predicted behavior and the continuum simulations) arises due to effects of the boundaries on the continuum simulations.
We use $\mathbf{u} = 0$ boundary conditions for the flow field in the confined continuum simulations, but Eq.~\eqref{eqn:PlusPlusVelocity} corresponds to defects in an infinite, not bounded, domain.
These boundary conditions reduce the flow magnitude and spatial extent, and, hence, increase the critical distance between defects.
The vertical shift factor, here found to be $\approx 1.2$, likely depends on both $\ell_h$ and the size of the computational domain, though a comprehensive study of this effect is outside the scope of the present work.
Additionally, because defects can spontaneously nucleate for higher activities in the continuum simulations, we cannot probe the high activity scaling result.
In the lower activity regime, however,
we obtain good qualitative agreement between our analytic model and
the behavior of defects in continuum simulations and experiments.

\subsection{Two $-1/2$ Defects}

The velocity of a $-1/2$ defect in the presence of another $-1/2$ defect is
\begin{multline} \label{eqn:MinusMinusVelocity}
    \mathbf{v}^{--} = \left[\frac{1}{r} - \frac{\alpha}{16 \ell_h^2}\chi_1\left(\frac{r}{2\ell_h}\right)\right]\mathbf{\hat{r}} \\ + \frac{\alpha}{4 \ell_h}\left[\beta_3\left(\frac{r}{2\ell_h}\right)
    \mathbf{\hat{c}} + \beta_4\left(\frac{r}{2\ell_h}\right)\mathbf{\hat{d}}\right] - \frac{\alpha}{16 \ell_h^2}\chi_3\left(\frac{r}{2\ell_h}\right) \mathbf{\hat{f}}
\end{multline}
where \textcolor{black}{$\mathbf{\hat{c}} = \left[\cos(3\theta - 2\varphi_0),-\sin(3\theta - 2\varphi_0)\right]$, $\mathbf{\hat{d}} = \left[\cos(5\theta - 2\varphi_0),\sin(5\theta - 2\varphi_0)\right]$, and $\mathbf{\hat{f}} = \left[\cos(7\theta - 4\varphi_0),-\sin(7\theta - 4\varphi_0)\right]$}. 
As with the two $+1/2$ defects, it is instructive to decompose the velocity into components parallel and transverse to the direction between the defects:
\begin{multline} \label{eqn:MinusMinusPar}
    \textcolor{black}{v_{||}^- = \frac{1}{r} - \frac{\alpha}{16\ell_h^2}\chi_1} \\
    \textcolor{black}{+ \frac{\alpha}{4\ell_h}\left(\beta_3 + \beta_4\right)\cos(4\theta - 2\varphi_0) - \frac{\alpha}{16\ell_h^2}\chi_3\cos(8\theta - 4\varphi_0)}
\end{multline}
\begin{multline} \label{eqn:MinusMinusPerp}
    \textcolor{black}{v_{\perp}^- = -\frac{\alpha}{4\ell_h}\left(\beta_3 - \beta_4\right)\sin(4\theta - 2\varphi_0)} \\
    \textcolor{black}{+ \frac{\alpha}{16\ell_h^2}\chi_3\sin(8\theta - 4\varphi_0).}
\end{multline}

\begin{figure}
\centering
    \includegraphics[width = \columnwidth]{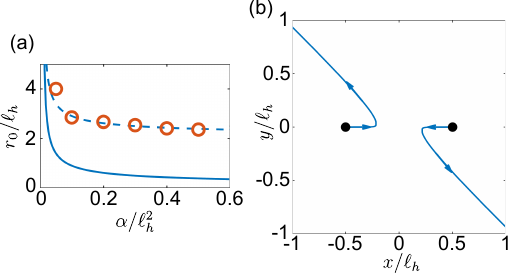}
\caption{(a) Scaled critical distance $r_0/\ell_h$ at which the velocity of $-1/2$ defects $\mathbf{v} = 0$ for $4\theta - 2\varphi_0 = 0$ versus $\alpha/\ell_h^2$.
\textcolor{black}{The solid line shows the calculated $r_0/\ell_h$ from Eq.~\eqref{eqn:MinusMinusPar}.}
\textcolor{black}{Orange points show} $r_0/\ell_h$ versus $\alpha/\ell_h^2$ for two $-1/2$ defects in confined continuum simulations with $\ell_h = 10$ and domain radius $R = 30$. 
\textcolor{black}{The dashed curve is a reproduction of the solid curve with a vertical shift of $2$.}
(b) Example trajectories of two $-1/2$ defects with initial $x = -0.5 \ell_h$ and $x = 0.5\ell_h$ at $\varphi_0 = 0.001$.}
    \label{fig:MinusMinusVelocity}
\end{figure}

Equation \eqref{eqn:MinusMinusPerp} shows that the transverse component of the velocity will be zero if $4\theta - 2\varphi_0 = n\pi$. 
In this case, two $-1/2$ defects can only move toward or away from one another. 
From Eq.~\eqref{eqn:MinusMinusPar}, we find that if $4\theta - 2\varphi_0 = (2n + 1)\pi$, the defects repel each other.  
Geometrically, this occurs when two $-1/2$ defects ``point'' towards one another if viewed as having a triangular shape. 
On the other hand, when the defects point away from one another for $4\theta - 2\varphi = 2n\pi$, the hydrodynamic interactions compete with the Coulomb force on the defects and there is a critical distance $r_0$ at which $v_{||} = 0$. 
In Fig.~\ref{fig:MinusMinusVelocity}(a) we plot this critical distance as a function of $\alpha/\ell_h^2$ for $\ell_h = 10$. 
For two $-1/2$ defects, $r_0$ rapidly decreases as $\alpha/\ell_h^2$ increases until $\alpha/\ell_h \approx 1$ and the value of $r_0$ saturates. 
As we did with the $+1/2$ defects, in Fig.~\ref{fig:MinusMinusVelocity}(a) we plot $r_0/\ell_h$ for several confined continuum simulations with two $-1/2$ defects. 
The dashed line in Fig.~\ref{fig:MinusMinusVelocity}(a) is the solid curve in the figure with a vertical shift, $r_0/\ell_h + 2$,
showing that apart from the vertical shift,
the
finite system continuum simulations behave qualitatively
similarly to the curve derived analytically for an infinite domain.
As with the $+1/2$ defects in the previous section, we suspect the quantitative difference to result from the confinement of the continuum simulations, where the flow speeds are reduced even more than in the case of two $+1/2$ defects.
We expect that the value of the vertical shift, found to be $2$ for the parameters used here, should be dependent on $\ell_h$ and the domain size of the confined system.

Although a critical distance between $-1/2$ defects exists, the velocity equations do not support stable, bound defect pairs.
\textcolor{black}{This is because the dynamics are explicitly dependent on $\theta$, and, in particular, $v_{\perp}$ changes as the defects rotate around one another.}
\textcolor{black}{Further, expanding around the fixed points of Eq.~\eqref{eqn:MinusMinusPerp} shows that $4\theta - 2\varphi_0 = 2n\pi$ are unstable fixed points, while $4\theta - 2\varphi_0 = (2n + 1)\pi$ are stable fixed points.}
\textcolor{black}{The model thus predicts that active $-1/2$ defects will tend to reorient toward a stable relative orientation when interacting.}
We illustrate this instability in Fig.~\ref{fig:MinusMinusVelocity}(b) by plotting the computed trajectories of two $-1/2$ defects with initial positions $(-0.5\ell_h,0)$ and $(0.5\ell_h,0)$, for $\varphi_0 = 0.001$, $\ell_h = 10$ and $\alpha/\ell_h^2 = 0.5$. 
Initially the defects move towards each other until they rotate enough for the interaction to become repulsive. 
They then repel each other along a line such that asymptotically approaches $4\theta - 2\varphi_0 = \pi$. 
This instability provides an explanation for why bound states of $-1/2$ defect pairs are not observed in continuum simulations and experiments.

\subsection{$\pm 1/2$ Defects}

Unlike the previous two cases we have examined, $\pm 1/2$ defect pairs interact non-reciprocally \cite{angheluta21}. 
That is, $\mathbf{v}^{+-} \neq -\mathbf{v}^{-+}$ and, thus, the predicted trajectories will not be symmetric. 
The velocities of the defects are given by
\begin{multline} \label{eqn:PlusMinusVelocity}
    \mathbf{v}^{+-} = \left[-\frac{1}{r} - \frac{\alpha}{16\ell_h^2}\chi_1\left(\frac{r}{2\ell_h}\right)\right]\mathbf{\hat{r}} - \frac{\pi\alpha}{4\ell_h}\mathbf{\hat{p}} \\ + \frac{\alpha}{4\ell_h}\left[\beta_3\left(\frac{r}{2\ell_h}\right)\mathbf{\hat{a}^{*}} + \beta_4\left(\frac{r}{2\ell_h}\right)\mathbf{\hat{c}^{*}}\right] - \frac{\alpha}{16\ell_h^2}\chi_3\left(\frac{r}{2\ell_h}\right)\mathbf{\hat{g}},
\end{multline}
\begin{multline} \label{eqn:MinusPlusVelocity}
    \mathbf{v}^{-+} = \left[-\frac{1}{r} - \frac{\alpha}{16\ell_h^2}\chi_1\left(\frac{r}{2\ell_h}\right)\right]\mathbf{\hat{r}} \\ + \frac{\alpha}{4\ell_h}\left[\beta_1\left(\frac{r}{2\ell_h}\right)\mathbf{\hat{a}^{*}} + \beta_2\left(\frac{r}{2\ell_h}\right)\mathbf{\hat{c}^{*}}\right] - \frac{\alpha}{16\ell_h^2}\chi_2\left(\frac{r}{2\ell_h}\right)\mathbf{\hat{h}}
\end{multline}
where \textcolor{black}{ $\mathbf{\hat{a}^{*}} = [\cos(\theta - 2\varphi_0),-\sin(\theta - 2\varphi_0)]$, $\mathbf{\hat{c}^{*}} = [\cos(3\theta - 2\varphi_0),\sin(3\theta - 2\varphi_0)]$, $\mathbf{\hat{g}} = [\cos(5\theta - 4\varphi_0),\sin(5\theta - 4\varphi_0)]$, $\mathbf{\hat{h}} = [\cos(3\theta - 4\varphi_0),-\sin(3\theta - 4\varphi_0)]$}, and we note that we have written the velocities in coordinates where the origin is located at the other defect.

\begin{figure}
\centering
    \includegraphics[width = \columnwidth]{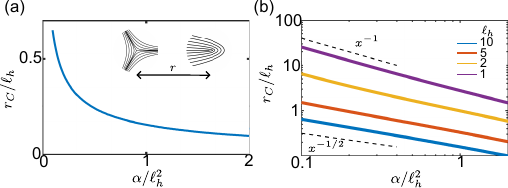}
\caption{(a) Scaled critical unbinding distance $r_C/\ell_h$ versus $\alpha / \ell_h^2$ for $\ell_h = 10$ and $\varphi_0 = \pi/2$. The inset shows a schematic of the orientation of the defects. (b) $r_C/\ell_h$ versus $\alpha /\ell_h^2$ for various screening lengths $\ell_h$ plotted on a log-log scale. The lower dashed line indicates $x^{-1/2}$ scaling while the upper dashed line indicates $x^{-1}$ scaling.}
    \label{fig:PlusMinusUnbind}
\end{figure}

We first analyze the interaction by examining the limiting case
where the global phase is $\varphi_0 = \pi/2$ and the axes are oriented so that $\theta = 0$
\textcolor{black}{for the $+1/2$ defect and $\theta = \pi$ for the $-1/2$ defect.} 
In this case the comet head of the $+1/2$ defect is pointing ``away'' from the $-1/2$ defect [see Fig.~\ref{fig:PlusMinusUnbind}(a)]. 
The velocity difference, and, hence, the rate at which the distance between defects changes, is
\begin{equation} \label{eqn:PlusMinusDistance}
    \textcolor{black}{\frac{d r}{d t} = -\frac{2}{r} + \frac{\alpha}{4\ell_h}\left(\pi - \beta_1 + \beta_4\right) - \frac{\alpha}{16 \ell_h^2}\left(2\chi_1 + \chi_2 + \chi_3\right).}
\end{equation}
Given a value for $\alpha$ and $\ell_h$, Eq. \eqref{eqn:PlusMinusDistance} predicts that there will be a critical distance $r_C$ above which the defects will move away from one another and below which the defects will inevitably annihilate. 
We argue that in a large active nematic system with many defects, the average distance between defects $\langle r_d \rangle \sim r_C$ since this is the distance at which $+1/2$ defects may unbind from $-1/2$ defects. 

In Fig.~\ref{fig:PlusMinusUnbind}(a) we plot $r_C/\ell_h$ versus $\alpha/\ell_h^2$ for $\ell_h = 10$. Here $r_C \to \infty$ as $\alpha \to 0$ as expected, since passive $\pm1/2$ defect pairs will always annihilate.
As shown by the log-log plot in Fig.~\ref{fig:PlusMinusUnbind}(b),
we find that $r_C \sim \alpha^{-1/2}$,
a scaling that has been observed for $\langle r_d \rangle$ in continuum simulations \cite{giomi15,Hemingway16}. 
In some sense, this scaling is expected because the active length scale 
\textcolor{black}{(the length scale that arises due to the competition between active and elastic forces \cite{doo18})}
$\ell_a \propto \alpha^{-1/2}$; however, a naive approximation would lead to the expectation that $dr/dt \sim v - 1/r$, where $v$ is the speed of the $+1/2$ defect, and hence that $r_C \sim \alpha^{-1}$. 
Only by including all of the hydrodynamic effects can the $\alpha^{-1/2}$ scaling be predicted. 
Interestingly, we find that the screening length $\ell_h$ plays an important role in this scaling. 
In Fig.~\ref{fig:PlusMinusUnbind}(b)
we also plot $r_C/\ell_h$ versus $\alpha/\ell_h^2$ for $\ell_h  \in \{1,2,5\}$. 
For small screening lengths and small activity, $r_C \sim \alpha^{-1}$, confirming the naive approximation. 
In this regime, the self propulsion is the dominant interaction term, and only at higher activities is the $\alpha^{-1/2}$ scaling restored. 
The crossover between regimes as a function of $\alpha$ occurs for higher relative activities at smaller screening lengths. 
Thus, we predict that active nematic experiments and simulations performed in the ``dry'' limit (smaller screening length) for relatively small activities will measure $\langle r_d \rangle \sim \alpha^{-1}$.

\textcolor{black}{To gain some analytical insight we use the approximation of $\beta_1$ in the limit of $r_C << \ell_h$ [see Fig. \ref{fig:VelocityFunctions}(a) and the Appendix].}
\textcolor{black}{Setting $\beta_1(r_C/2\ell_h) = (\pi^2/4)/(\pi/4 + r_C/2\ell_h)$ in Eq.~\eqref{eqn:PlusMinusDistance} and neglecting the terms which are small in this limit gives
\begin{equation}
    r_C \approx \frac{1 + \sqrt{1 + \ell_h^2(\pi^2/4)(\alpha/\ell_h^2)}}{\ell_h(\pi/4)(\alpha/\ell_h^2)}
\end{equation}
where we have written the right hand side explicitly in terms of $\alpha/\ell_h^2$ to match the scaling in Fig.~\ref{fig:PlusMinusUnbind}.
From this approximation, we see that if either $\ell_h$ or $\alpha$ is large then we recover $r_C \sim (\alpha/\ell_h^2)^{-1/2}$, while if  both $\ell_h$ and $\alpha$ are small $r_C \sim (\alpha/\ell_h^2)^{-1}$.}
\textcolor{black}{Thus, both limits are able to be captured by this approximation, and we emphasize that the full model can interpolate between these two regimes.}

\begin{figure}
\centering
    \includegraphics[width = \columnwidth]{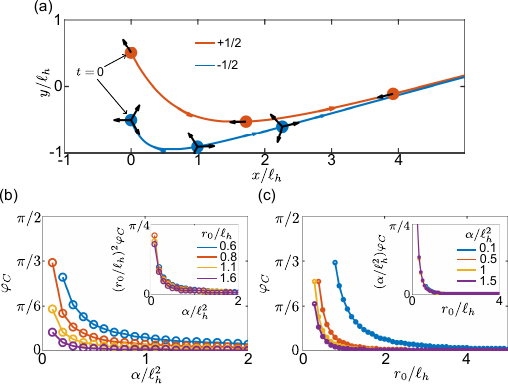}
    \caption{(a) Example trajectories of $\pm 1/2$ defects with initial positions $y^+ = 0.5 \ell_h$ and $y^- = -0.5 \ell_h$, $\varphi_0 = \pi/12$, and $\ell_h = 10$. The points are time snapshots of the positions of the defects at $t/\Delta t = 0$, $t/ \Delta t = 5000$ and $t / \Delta t = 10000$. The arrows on the points show the orientation of the defect. (b) Critical global phase $\varphi_C$ above which $+1/2$ defects deflect from $-1/2$ defects versus $\alpha / \ell_h^2$ for $\ell_h = 10$ at various initial defect separations $r_0/\ell_h$. Inset: the same data with the vertical axis scaled by $(r_0/\ell_h)^2$. (c) $\varphi_C$ versus $r_0/\ell_h$ for various $\alpha/\ell_h^2$. Inset: the same data with the vertical axis scaled by $\alpha/\ell_h^2$.}
    \label{fig:PlusMinusTrajectory}
\end{figure}

One might predict that if the comet head of the $+1/2$ defect is oriented toward the $-1/2$ defect, then annihilation will always occur. 
This is precisely true if $2\theta - 2\varphi_0 = 2n\pi$; however, due to the orientational coupling and self propulsion of the $+1/2$ defect, if $2\theta - 2\varphi_0 \neq 2n\pi$, the $+1/2$ defect may deflect away from the $-1/2$ defect. 
In this more general case, the complicated non-reciprocity of the $\pm 1/2$ defect velocities does not lend itself to a straightforward analytical analysis. 
Instead, we analyze the interaction by numerically integrating Eqs.~\eqref{eqn:PlusMinusVelocity} and $\eqref{eqn:MinusPlusVelocity}$ and observing the resulting trajectories. 
In Fig.~\ref{fig:PlusMinusTrajectory}(a) we plot an example trajectory with $\varphi_0 = \pi/12$. In this case, although the $\pm 1/2$ defect pair distance initially decreases, eventually it begins to grow with time, as can be seen by the three snapshot points plotted along the trajectory. 
We note that although $-1/2$ defects are typically thought of as ``passive'' particles, their motion is not only influenced by Coulomb interactions and can be pushed or pulled by the flows of other defects, which can be seen in the trajectory in Fig. \ref{fig:PlusMinusTrajectory}(a).

\begin{figure}
\centering
    \includegraphics[width = \columnwidth]{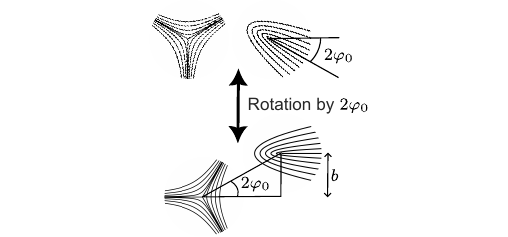}
\caption{Schematic depiction of rotating the system axes to define the impact parameter $b$ in terms of $\varphi_0$.}
    \label{fig:TwoDefectRotate}
\end{figure}

We find that, if defects are initially oriented along the $x$-axis, there is a critical global phase, $\varphi_C$, depending on initial separation and activity, above which $+1/2$ defects will deflect and not annihilate. 
To study this systematically, we compute trajectories of $\pm 1/2$ defect pairs with varying initial distances and orientations. 
We consider the two defects ``annihilated'' if they come within one nematic correlation length of each other. 
In our normalized units this occurs when $r < 1$. 
In Fig.~\ref{fig:PlusMinusTrajectory}(b) we plot the critical angle $\varphi_C$ versus $\alpha / \ell_h^2$ for
$\ell_h=10$ at various initial distances, and
in Fig.~\ref{fig:PlusMinusTrajectory}(c) we plot $\varphi_C$ versus $r_0/\ell_h$ for various activities. 
The insets of these figures show that $\varphi_C \sim r_0^{-2}\alpha^{-1}$, or $\varphi_C \sim (\ell_a / r_0)^2$.
We can also map this critical angle to a critical impact parameter $b_C$ by rotating our axes by $2\varphi_C$ (so that $\varphi_0 \to 0$). 
Then, $b_C = r_0\sin2\varphi_C$ so that $b_C \approx 2\ell_a^2/r_0$. 
We sketch the rotation leading to this mapping in Fig. \ref{fig:TwoDefectRotate}.

\textcolor{black}{Our analysis of the model's predictions for the critical unbinding lengths and angles of $\pm 1/2$ annihilation do not assume a particular method of nucleation. 
One future extension of the model would be to implement specific nucleation methods. 
However, once the defects have nucleated, our analysis should apply to situations where the $\pm 1/2$ pair are well separated or screened from other defects in the system. 
Therefore, we expect our analysis to apply to defects nucleated from the usual bend instability \cite{doo18}, as well as defects nucleated via recently observed disorder-order transitions in phototactic active nematics \cite{repula24}.}

\section{Multiple Defects in Circular Confinement}

We next consider the computation of the trajectories of many defects in a confined circular geometry.
Confining the space in which the particles are allowed to move better facilitates comparisons between particle-based simulations and experiments or continuum simulations. 
The model presented in Sec.~\ref{sec:DefectVelocity} could be used to simulate many defects in an infinite system. 
However, because $+1/2$ defects are able to self propel, these defects will eventually move off to infinity. 

To simulate a confined circular geometry, we use the method of images for defects in a circular domain. 
For each defect located at polar coordinate $(r,\theta)$, a defect of equal charge is placed at $(R^2/r,\theta)$, where $R$ is the radius of confinement. 
The corresponding lowest energy director angle is given by
\begin{multline} \label{eqn:CircConfineDirector}
    \phi(x,y) = \frac{1}{2}\sum_{i_+}\left[\arctan\left(\frac{y - r_{i_+}\sin\theta_{i_+}}{x - r_{i_+}\cos\theta_{i_+}}\right) \right. \\ \left. + \arctan\left(\frac{y - R^2/r_{i_+}\sin\theta_{i_+}}{x - R^2/r_{i_+}\cos\theta_{i_+}}\right) - \theta_{i_+}\right] \\ - \frac{1}{2}\sum_{i_-}\left[\arctan\left(\frac{y - r_{i_-}\sin\theta_{i_-}}{x - r_{i_-}\cos\theta_{i_-}}\right) \right. \\ + \left. \arctan\left(\frac{y - R^2/r_{i_-}\sin\theta_{i_-}}{x - R^2/r_{i_-}\cos\theta_{i_-}}\right) - \theta_{i_-}\right] + \varphi_0
\end{multline}
where the polar angles of each defect $\theta_i$ are subtracted so that $\varphi_0$ gives the director angle offset at $R$. 
Equation \eqref{eqn:CircConfineDirector} describes a director field which will always have a fixed director at $R$. 
If the total topological charge of the system is zero, then $\phi(r = R) = \varphi_0$, a constant. 
The total topological charge does not have to be zero, however. 
Equation \eqref{eqn:CircConfineDirector} will give the lowest energy director regardless of the total topological charge. 
We can then obtain the orientation of each defect using $\varphi_j = \lim_{\epsilon\to 0} \phi(x_j + \epsilon,y_j)$.

\begin{figure*}
\centering
    \includegraphics[width = \textwidth]{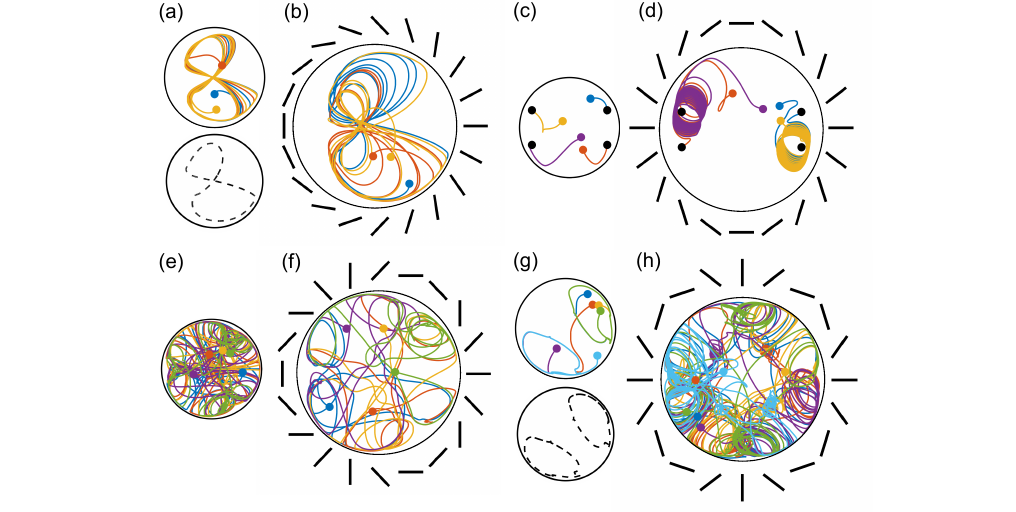}
\caption{(a) Example trajectories given by the analytical model for three $+1/2$ defects in a confined system with radius $R = 3$ and $\alpha = 2$. Different colors depict different defects. The dashed line in the lower figure depicts the late-time stable trajectory followed by all three particles. (b) Example trajectories for three $+1/2$ defects with $R = 5$ and $\alpha = 2$. The lines outside the circle depict the director field at the boundary, which also apply to the system in (a). (c) Example trajectories for four $+1/2$ defects with $R = 3$ and $\alpha = 2$. The black dots indicate the stable final positions of the defects. (d) Example trajectories for four $+1/2$ defects with $R = 5$ and $\alpha = 2$. (e) Example trajectories for five $+1/2$ defects with $R = 3$ and $\alpha = 3$. No stable points or trajectories were observed for the number of integration steps we considered. (f) Example trajectories for five $+1/2$ defects with $R = 5$ and $\alpha = 3$. (g) Example trajectories for six $+1/2$ defects with $R = 3$ and $\alpha = 3$. The dashed lines in the lower figure depict the late-time stable trajectories followed by three particles each. (h) Example trajectories for six $+1/2$ defects with $R = 5$ and $\alpha = 3$.}
    \label{fig:ConfinedParticles}
\end{figure*}

In confined systems, it is typical to assume no-slip boundary conditions for the fluid velocity. 
However, to our knowledge, no analytical solution to Eq. \eqref{eqn:Stokes} is available for $\mathbf{u}(R) = 0$ boundary conditions. 
Thus, here we neglect the effect of the boundaries on the flow solution, and limit our considerations to the case of small screening length, namely $\ell_h = 1$. 
In this parameter regime, it was shown in Ref.~\cite{ronning22} that the flow solutions for a single defect are only weakly perturbed by no slip boundary conditions when the analytical solution for an infinite domain was compared to bounded numerical solutions. 
Finally, we assume that defects in the system do not feel the flow solutions from the image charges but experience only the Coulomb force from elastic forces. 
For the parameters that we explore, this assumption only results in small quantitative changes to the defect motion.
In what follows, we integrate the equations of motion for each defect using a forward Euler method with time step $\Delta t = 0.001$ and initialize the system with randomly generated positions of the defects.

We first examine multiple $+1/2$ defects in confinement. 
For two $+1/2$ defects, the results of Sec.~\ref{sec:TwoDefects} apply and the defects will rotate around one another if $\varphi_0 \neq 0,\pi/2$. 
If $\varphi_0 = 0, \pi/2$, the defects will find stable positions with $\mathbf{v} = 0$. 
For three $+1/2$ defects, we find that the defects tend to braid one another in a manner similar to that seen in recent experiments \cite{Memarian24}. 
In Fig.~\ref{fig:ConfinedParticles}(a,b) we show example trajectories for three defects with system radii $R = 3$ and $R = 5$, respectively. 
For the smaller system size, the defects maintain tighter ``figure eight''  trajectories, as seen in the experiments, until eventually settling into the stable figure eight trajectory shown by the dashed line in Fig.~\ref{fig:ConfinedParticles}(a). For the larger system size the defects do not follow perfect braiding trajectories, but instead traverse figure eight pathways bounded by outer and inner figure eight loops. Here we do not observe the same stabilization of the trajectories that appears in the more confined system.
Movies of the integrated analytical model are available in
the Supplemental Material (1,2). 
The ability of our model to predict the braiding behavior seen in experiments highlights the importance of orientational coupling in the active nematic. 
Such behavior could not be reproduced with a simpler active Brownian particle model where only Coulomb interactions are present.

We also examine trajectories for four, five, and six defects. 
We plot example trajectories for all of these cases in
Fig.~\ref{fig:ConfinedParticles}(c-h)
and show movies of their motion in the Supplemental Material (3--8). 
For four $+1/2$ defects we find that there is a stable configuration where the defects sit in a rectangular formation and $\mathbf{v} = 0$ for each defect.
At small confinement the system rapidly relaxes to this configuration
regardless of the random initial positions,
as shown in Fig.~\ref{fig:ConfinedParticles}(c). 
For larger confinement, the defects eventually find the stable configuration, but only after a long transient motion of two bound states of two $+1/2$ defects that rotate around one each other, as illustrated in Fig.~\ref{fig:ConfinedParticles}(d). 
We find that stable configurations exist for four $+1/2$ defects for all values of $\varphi_0$. 
We note that, unlike the case for two $+1/2$ defects, the value of $\varphi_0$ has no influence over the existence of a stable state. 
This is because a total topological charge of $m = 1$ is a special case since it is the only arrangement for which the relative angle of the director with respect to the boundary is constant. 
For any other total topological charge, the director angle relative to the boundary changes continuously, and hence any rotation of the director rotates the overall configuration.
To emphasize this, in Figs.~\ref{fig:ConfinedParticles}(b,d,f,h) we indicate the director boundary condition.
We also note that although the flow solutions we assume here do not respect no-slip boundary conditions, the symmetry of the configuration dictates that a qualitatively similar configuration should still exist for the case of no-slip conditions.

For five defects we no longer find stable defect configurations. 
In Figs.~\ref{fig:ConfinedParticles}(e,f) we plot example trajectories for five $+1/2$ defects and do not observe any closed defect trajectories, meaning that there is not a stable attractor for either system size. 
For six defects, when $R = 3$ we observe two periodic trajectories that contain three defects each, as shown in 
Fig.~\ref{fig:ConfinedParticles}(g). 
The periodic trajectories may change their orientation for different random initial conditions or values of $\varphi_0$.
Upon increasing the system size to $R = 5$, however, we no longer observe stable periodic trajectories, as illustrated in Fig.~\ref{fig:ConfinedParticles}(h). 
We note that while it is beyond our scope here to extensively study the effect of $\alpha$ on these trajectories, we find that the qualitative results do not change upon a cursory survey of varying activities. 
That is,
only the quantitative details of the points and trajectories vary when small
changes are made to the activity, but
the existence of stable points or trajectories is unaffected.

\begin{figure}
\centering
    \includegraphics[width = \columnwidth]{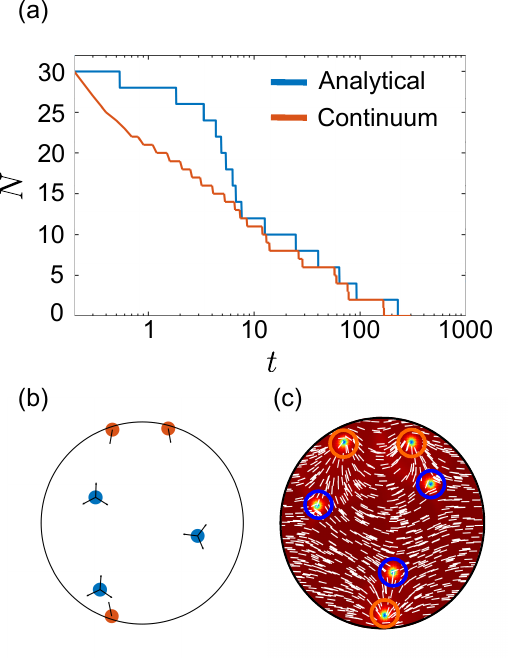}
\caption{(a) Comparison of the number of defects $N$ versus time $t$ for the numerically integrated analytical model and a continuum simulation with $R = 50$, $\alpha = 2$, $\ell_h = 1$, $k = 1$, and $\varphi_0 = 0$. The analytical model was initialized with random positions of 15 $+1/2$ and 15 $-1/2$ defects, while the continuum simulation was initialized with a director field that has the same random positions of defects used in the analytical model. Time is given in the dimensionless units outlined in Sec.~\ref{sec:DefectVelocity}. (b) Late-time configuration snapshot of the numerically integrated analytical model. (c) Late-time configuration snapshot of the continuum model. $+1/2$ defects are circled in orange while $-1/2$ defects are circled in blue.}
    \label{fig:CoarsenExample}
\end{figure}

Finally, we compare active nematic defect coarsening
in the analytical model and a continuum simulation. 
For the analytical model, we initialize a system of radius $R = 50$ with 15 $+1/2$ defects and 15 $-1/2$ defects at random positions. 
We then numerically integrate their velocities and remove $\pm 1/2$ pairs that come within $r < 1$ of each other. 
For the continuum simulation, we initialize the system with a director field that has defects at the same positions as the analytical model. For both computations, we use $\alpha/\ell_h^2 = 2$, $\ell_h = 1$, $k =1$, and $\varphi_0 = 0$. 
Importantly, for these parameters, the continuum simulation does not spontaneously nucleate defects.
In Fig.~\ref{fig:CoarsenExample}(a) we plot the number of defects $N$ versus dimensionless simulation time $t$ for both the computations with logarithmic scaling on the $t$-axis. 
At early times, when the system is more dense, the coarsening dynamics are different, with faster annihilation of defects in the continuum simulation. 
However, at longer time scales, the two curves have roughly logarithmic scaling with time. 
We also find that at late times, $+1/2$ defects move out toward the boundaries while $-1/2$ defects are left in the bulk. The $+1/2$ defects congregate at polar angles $\theta = \varphi_0 \pm \pi/2$ due to the orientational coupling between the defects and director field.
In Fig.~\ref{fig:CoarsenExample}(b) we plot a late-time snapshot of an analytically integrated computation while in Fig.~\ref{fig:CoarsenExample}(c) we plot a late-time snapshot of a continuum configuration. 
In both cases the charge separation occurs while the defects continue to coarsen. 
While a full understanding of confined active nematic defect coarsening would require many more realizations and an extensive study of the effect of model parameters,
this example highlights a potential future application of the analytical model, which requires far less computational time and resources to simulate than the continuum model.

\section{Discussion}

As we have demonstrated, our analytical model qualitatively reproduces features of active nematic defect interactions observed in both continuum simulations and experiments.
\textcolor{black}{It is also able to reproduce predictions from previous analytical models of defect motion, such as the mutual rotation of two $+1/2$ defects, and the deflection of $+1/2$ defects by $-1/2$ defects \cite{angheluta21,Vafa22}.}
\textcolor{black}{By including a well-defined screening length, the model can interpolate between wet and dry regimes of active nematics.}
\textcolor{black}{Additionally, because we employ a linear perturbation approximation, the model may be used to investigate the motion of multiple defects interacting with one another.}

One of the primary advantages of the analytical model is the ease with which the effect of different parameters may be investigated.
For example, in future studies, the model can be used to probe
the effect of screening length and tumbling parameter.
Another advantage is the modular nature of the theory. 
One may, for example, investigate the effect of elastic interactions by changing the coefficient for the Coulomb force term; or one could impose different flow profiles by adding the appropriate terms to the velocity.
\textcolor{black}{Since the velocity of defects is derived explicitly from topological charge conservation via the methods of Halperin and Mazenko\cite{halperin81,liu92,mazenko97,mazenko99}, we do not need to interpret the results in terms of auxiliary fields\cite{Cortese18,ZhangYi20}.}
\textcolor{black}{Having an explicit form for the defect velocities is particularly amenable for computational implementation.}

Our analytical model achieves its greatest accuracy under the separation of length scales $a < \langle r_d \rangle < \ell_a$, where $a$ is the core size, $\langle r_d \rangle$ is the average separation of defects, and $\ell_a$ is the active length scale. 
Our linear perturbation approximation, which allows us to treat the flows of individual defects separately, breaks down when the cores of defects overlap. 
Additionally, we have neglected the effect of topologically trivial structures such as bend walls,
which also generate flows in the system \cite{Head24}, 
on the motion of defects.
When $\langle r_d \rangle > a$, the effect of defect core overlap will be minimal, and when $\ell_a > \langle r_d \rangle$, structures such as bend walls will be quickly relaxed due to elastic forces. 

We make the
approximation that the director field is always given by the minimal energy configuration. 
It has been shown in experiments that this is not always the case for active nematics \cite{pearce21,Pearce21b}, and induced torques, transverse velocities, and \textcolor{black}{orientational noise} may play an important role in the motion of active nematic defects \cite{vromans16,tang17,shankar18,Shankar19}.
This presents an opportunity for a future extension of the analytical model, since it has been shown that the defect velocity formalism used here can capture the transverse motion induced by these torques \cite{schimming23}.
However, to our knowledge, the flows induced by the relative twisting of defects in active nematics are not yet well understood. 
If approximate equations for this can be derived, then the orientation of topological defects in active nematics may be treated independently of their positions,
\textcolor{black}{while still taking into account the intrinsic coupling of the director field and defect orientation,}
and a true particle model could be formulated
\textcolor{black}{using the present theoretical framework.} 
Accurately treating the orientation of defects independently would allow for many additional features to be included in the model such as defect nucleation or orientational noise. It would also be possible to include periodic boundary conditions, which cannot be treated in the present model due to instantaneous jumps in the orientation when defects cross the periodic boundary. 
In each case, it would be possible to maintain the orientational coupling that we have shown is important when describing active nematic defect motion.

An interesting extension of the methodology presented here would be to treat
line defects in three-dimensional active nematics. 
Although it has been shown that the defect velocity
formalism we employ here can be generalized to line defects in three dimensions \cite{schimming23,Zushi24}, there has been much less work in understanding defect dynamics in three dimensional active nematics both computationally \cite{Copar19,kralj23}, and, especially, experimentally \cite{duclos20}. 
Recent theoretical work has described some aspects of defect loop motion \cite{binysh20,Houston22}, and it will be interesting to compare and combine the results of the methodologies used here to obtain a clearer picture of defect motion in active nematics.

\section{Summary}
We have presented an analytical model for active nematic defect motion based on recently derived theoretical techniques and solutions to the Stokes equation for active nematic defects. 
\textcolor{black}{Our work builds upon previous efforts to understand defect motion by applying a linear perturbation approximation to the Halperin-Mazenko formalism of defect dynamics.}
\textcolor{black}{Using this, explicit velocities for an arbitrary number of defects may be predicted.}
The model incorporates both the flow-induced long range interactions between defects and also the crucial orientational coupling between the defects.
We demonstrate that our model provides analytical predictions for the stable interaction radius adopted by bound states of pairs of $+1/2$ defects, the effective attraction that appears between pairs of $-1/2$ defects, and the way in which the critical unbinding length between a $+1/2$ and $-1/2$ defect pair scales with activity. The model can also capture the dependence of the scattering trajectory of a $+1/2$ defect from a $-1/2$ defect on activity. It can be applied to the braiding motion of multiple confined $+1/2$ defects as well as the coarsening behavior that occurs for a mixture of equal numbers of $+1/2$ and $-1/2$ defects under confinement.
It will serve as a useful tool for future studies.

\section*{Appendix}

\subsection*{Defect Velocity Calculations} \label{app:VelocityCalculations}

Here we present more detailed calculations leading to the expressions in Eqs.~\eqref{eqn:CoulombVelocity}, \eqref{eqn:AdvectionVelocity}, and \eqref{eqn:ShearVelocity}. For completeness, we start by briefly reviewing the arguments leading to Eq.~\eqref{eqn:DefectVelocity}, though more detailed derivations may be found in Refs.~\cite{angheluta21,schimming23} for nematic defects \textcolor{black}{and Refs.~\cite{liu92,mazenko97,mazenko99} for more general systems of defects.}

The topological defect density in two dimensions may be written as
\begin{equation} \label{eqn:DefectDensity}
    \rho(\mathbf{r},t) = \sum_i m_i \delta\left[\mathbf{r} - \mathbf{r_i}(t)\right]
\end{equation}
where $m_i$ is the winding number of the $i$th defect. Using the fact that zeros of $\mathbf{Q}$ correspond to topological defects in the nematic phase, the Dirac delta function may be rewritten over $\mathbf{Q}(\mathbf{r})$ and $\rho$ may be written as
\begin{align}
    \rho(\mathbf{r},t) &= D(\mathbf{r},t)\delta\left[\mathbf{Q}(\mathbf{r},t)\right] \\
    D &= \varepsilon_{k\ell}\varepsilon_{\mu \nu} \partial_k Q_{\mu \alpha} \partial_{\ell} Q_{\nu \alpha} \label{eqn:D}
\end{align}
where $D$ is related to the Jacobian transformation from real space to order parameter ($\mathbf{Q}$) space. Taking a time derivative of $D$ reveals
\begin{equation}
    \partial_t D = \partial_k \left( 2\varepsilon_{k\ell}\varepsilon_{\mu\nu}\partial_t Q_{\mu \alpha} \partial_{\ell} Q_{\nu \alpha}\right) \equiv \nabla \cdot \mathbf{J}_{D}.
\end{equation}
Thus, $D$ is a conserved field with current $\mathbf{J}_D$. Taking a time derivative of both expressions for $\rho$, rearranging terms, and enforcing delta function relations yields Eq.~\eqref{eqn:DefectVelocity}.

To get analytical approximations for the velocity, we must approximate $\mathbf{Q}$ near defect cores. We assume a linear core structure \cite{long21}:
\begin{equation} \label{eqn:QApprox}
    \mathbf{Q}(x,y) \approx \frac{x}{2}\left[\mathbf{\tilde{n}}_0 \otimes \mathbf{\tilde{n}}_0 - \mathbf{\tilde{n}}_1 \otimes \mathbf{\tilde{n}}_1\right]
    + \frac{y}{2}\left[\mathbf{\tilde{n}}_0 \otimes \mathbf{\tilde{n}}_1 + \mathbf{\tilde{n}}_1 \otimes \mathbf{\tilde{n}}_0\right]
\end{equation}
with $\mathbf{\hat{n}}_0 = \left(\cos\varphi_0,\sin\varphi_0\right)$, $\mathbf{\hat{n}}_1 = \mathbf{\hat{n}}_0 \cdot \bm{\varepsilon}$, and $\mathbf{\tilde{n}}_i = \mathbf{\hat{n}}_i + \tilde{\varphi} \left(\mathbf{\hat{n}}_i \cdot \bm{\varepsilon}\right)$. $\tilde{\varphi}$ is the director angle due to all other defects:
\begin{equation}
    \tilde{\varphi} = \sum_{j\neq i} m_j \arctan\left(\frac{y - y_j}{x - x_j}\right) + \varphi_0.
\end{equation}
Thus, we approximate the director distortion near the defect core of interest as a linear perturbation from all other defects.

The Coulomb velocity in Eq.~\eqref{eqn:CoulombVelocity} may be derived by substituting $\partial_t Q_{\mu \alpha} = -\delta F/\delta Q_{\mu \alpha} = (1/k)\nabla^2 Q_{\mu \alpha}$ into Eq.~\eqref{eqn:DefectVelocity}. 
Using Eq.~\eqref{eqn:QApprox} to compute the spatial derivative and setting $x = y = 0$ gives Eq.~\eqref{eqn:CoulombVelocity}. 
Given a fluid flow $\mathbf{u}(\mathbf{r} = 0)$, substituting $\partial_t Q_{\mu \alpha} = -(\mathbf{u}\cdot \nabla)Q_{\mu \alpha}$ into Eq.~\eqref{eqn:DefectVelocity} gives the advection component of the velocity, Eq.~\eqref{eqn:AdvectionVelocity}, after using Eq.~\eqref{eqn:QApprox} to compute the gradient of $\mathbf{Q}$. 
To get the shear component of the defect velocity, Eq.~\eqref{eqn:ShearVelocity}, we first note that upon substituting $\partial_t Q_{\mu \alpha} = S_{\mu \alpha}$ into Eq.~\eqref{eqn:DefectVelocity}, most of the terms are zero when $x = y = 0$ since $\mathbf{Q} = 0$ at the core of the defect. The only term that contributes to the velocity is $\partial_t Q_{\mu \alpha} = \lambda E_{\mu \alpha}$, and substitution of this into Eq.~\eqref{eqn:DefectVelocity} yields Eq.~\eqref{eqn:ShearVelocity}.

\subsection*{Radial flow solutions} \label{app:FlowSolutions}

Here we give the functional form of the distance dependent part of the flow and shear flow vectors for a single $\pm 1/2$ defect. The flow solutions are equivalent to those derived in Ref. \cite{ronning22}, though we have rewritten them to decompose them into radial and angular parts. The radial parts written here are infinite series which are tabulated to find the results in the main text.

The radial parts of the flow solutions are given by
\begin{align*}
    \beta_1(x) &= \sum_{k=0}^{\infty}\frac{x^{2k}}{(k!)^2}\left[\pi - (4k+3) x N_1(k)\right] \\
    N_1(k) &= \sum_{n=0}^{\infty}\left[\frac{(2n-1)!!}{(2n)!!}\right]^2\left[\frac{4n+1}{(n+k+1)^2(2n-2k-1)^2}\right] \\
    \beta_2(x) &= \sum_{k=0}^{\infty} \frac{x^{2k+1}}{(k!)^2}\left[\frac{-x\pi}{(k+2)(k+1)} + N_2(k)\right] \\
    N_2(k) &= \sum_{n=0}^{\infty}\left[\frac{(2n-1)!!}{(2n)!!}\right]^2 \\ &\times \left[\frac{\left[(2n-1)(4k+1)(n+1) - 4k^2\right](4n+1)}{(n+k+1)^2(2n-2k-1)^2(n+1)(2n-1)}\right] \\
    \beta_3(x) &= -\beta_2(x) \\
    \beta_4(x) &= \sum_{k=0}^{\infty} \frac{x^{2k+1}}{(k!)^2}\left[\frac{- x^3 \pi k!}{(k+4)!} + N_4(k)\right] \\
    N_4(k) &= \sum_{n=0}^{\infty}\left[\frac{(2n-1)!!}{(2n)!!}\right]^2 \\ &\times \left[\frac{4n(n-1)}{(n+1)(n+2)(2n-2k-1)^2} \right. \\ &\qquad \left. - \frac{(2n+1)(2n+3)}{(2n-1)(2n-3)(n+k+1)^2}\right].
\end{align*}
The radial parts of the strain rate solutions are given by
\begin{align*}
    \chi_1(x) &= -3N_1(0) + \sum_{k=1}^{\infty} \frac{x^{2k-1}}{(k!)^2}\left[2\pi k - (2k+1)(4k+3)x N_1(k)\right] \\
    \chi_2(x) &= \sum_{k=0}^{\infty} \frac{x^{2k}}{(k!)^2}\left[\frac{-2\pi k x}{(k+2)(k+1)} + (2k-1)N_2(k)\right] \\
    \chi_3(x) &= \sum_{k=0}^{\infty} \frac{x^{2k}}{(k!)^2}\left[\frac{-2\pi k x^3 k!}{(k+4)!} + (2k-3)N_4(k)\right].
\end{align*}

\textcolor{black}{To obtain the small $x$ approximation of $\beta_1$ we compute the first two terms giving $\beta_1(x) = \pi - 4x + \ldots$}
\textcolor{black}{Comparing to a function of the form $f(x) = B/(A + x)$, we may write (for $x < 1$) $f(x) = B/A - (B/A^2)x + \ldots$}
\textcolor{black}{Thus, assuming $\beta_1(x)$ scales as $f(x)$ for small $x$, we identify $B = \pi^2/4$ and $A = \pi/4$, or $\beta_1(x) \approx (\pi^2/4)/(\pi/4 + x)$.}
\textcolor{black}{The inset of Fig.~\ref{fig:VelocityFunctions}(a) shows empirically that this is a reasonable approximation for small $x$.}

\subsection*{Continuum simulation details} \label{app:Continuum}

The continuum simulations shown in the main text solve Eqs.~\eqref{eqn:QEvo} and~\eqref{eqn:Stokes} for the $\mathbf{Q}$ tensor and flow velocity, $\mathbf{u}$, which are rendered dimensionless via the same scaling described in the main text. For the free energy, we use the Landau-de Gennes bulk free energy with a one-constant elastic term:
\begin{equation}
    F = \int\left[A|\mathbf{Q}|^2 + |\mathbf{Q}|^4 + \frac{1}{k}|\nabla\mathbf{Q}|^2\right] \, d\mathbf{r}.
\end{equation}
For all simulations we set $A = -0.5$, so at equilibrium the liquid crystal is in the nematic phase with $S_N = \sqrt{2}$. To numerically solve the dynamical equations, Eq.~\eqref{eqn:QEvo} is discretized in time with an implicit backward Euler scheme and a Newton-Raphson sub-iteration to solve the resulting nonlinear equations. Both Eqs.~\eqref{eqn:QEvo} and \eqref{eqn:Stokes} are discretized in space on either a square domain or a circular domain and we solve the weak form of partial differential equations using the MATLAB/C++ package FELICITY \cite{walker18}. At each time step, given a configuration of $\mathbf{Q}$ we first solve Eq.~\eqref{eqn:Stokes} for $\mathbf{u}$. We then use this solution to compute the next time step for $\mathbf{Q}$. For all simulations we use a time-step of $\Delta t = 0.1$. 

For the simulation snapshots shown in Fig.~\ref{fig:PlusPlusContinuum}(a), we use periodic boundary conditions on a square domain. Periodic boundary conditions necessitate that the total topological charge remains neutral. On the other hand, for the continuum simulation snapshots shown in Fig.~\ref{fig:PlusPlusContinuum}(b) and data shown in Figs.~\ref{fig:PlusPlusContinuum}(c) and \ref{fig:MinusMinusVelocity}(b), we use Dirichlet boundary conditions on a circular domain, with $\mathbf{u} = 0$, $S = S_N$, and varying director depending on the total topological charge. To induce two $+1/2$ defects, the angle of the director field at the boundary is given by $\phi(R) = \arctan(y/x) + \varphi_0$, while for two $-1/2$ defects it is given by $\phi(R) = -\arctan(y/x) + \varphi_0$.

For the continuum simulation where we compare defect coarsening, shown in Fig. \ref{fig:CoarsenExample}, we use Dirichlet boundary conditions with $\phi(R) = 0$, requiring a total topological charge of zero. To find the number of defects as a function of time, we integrate the magnitude of $D$, given in Eq.~\eqref{eqn:D} \cite{blow14,schimming22}:
\begin{equation*}
    N = \frac{1}{\pi S_N^2}\int |D| \, d\mathbf{r}.
\end{equation*}

\subsection*{Supplementary Movie Descriptions}

In all Supplemental Movies the points are the positions of defects, the arrows are the defect orientations $\mathbf{\hat{p}}$, and the black circles indicate the domain boundaries. Below we list the specific parameters for each movie.
\begin{itemize}
    \item Supplemental Movie 1: Three $+1/2$ defects, $R = 3$, $\alpha = 2$, $\ell_h = 1$.
    \item Supplemental Movie 2: Three $+1/2$ defects, $R = 5$, $\alpha = 2$, $\ell_h = 1$.
    \item Supplemental Movie 3: Four $+1/2$ defects, $R = 3$, $\alpha = 2$, $\ell_h = 1$.
    \item Supplemental Movie 4: Four $+1/2$ defects, $R = 5$, $\alpha = 2$, $\ell_h = 1$.
    \item Supplemental Movie 5: Five $+1/2$ defects, $R = 3$, $\alpha = 3$, $\ell_h = 1$.
    \item Supplemental Movie 6: Five $+1/2$ defects, $R = 5$, $\alpha = 3$, $\ell_h = 1$.
    \item Supplemental Movie 7: Six $+1/2$ defects, $R = 3$, $\alpha = 3$, $\ell_h = 1$.
    \item Supplemental Movie 8: Six $+1/2$ defects, $R = 5$, $\alpha = 3$, $\ell_h = 1$.
\end{itemize}

\section*{Conflicts of interest}

There are no conflicts to declare.

\section*{Acknowledgements}

We gratefully acknowledge the support of the U.S. Department of
Energy through the LANL/LDRD program for this work.
This work was supported by the U.S. Department of Energy through the Los Alamos National Laboratory. Los Alamos National Laboratory is operated by Triad National Security, LLC, for the National Nuclear Security Administration of the U.S. Department of Energy (Contract No. 89233218CNA000001).



\balance


\bibliography{LC} 

\providecommand*{\mcitethebibliography}{\thebibliography}
\csname @ifundefined\endcsname{endmcitethebibliography}
{\let\endmcitethebibliography\endthebibliography}{}
\begin{mcitethebibliography}{74}
\providecommand*{\natexlab}[1]{#1}
\providecommand*{\mciteSetBstSublistMode}[1]{}
\providecommand*{\mciteSetBstMaxWidthForm}[2]{}
\providecommand*{\mciteBstWouldAddEndPuncttrue}
  {\def\EndOfBibitem{\unskip.}}
\providecommand*{\mciteBstWouldAddEndPunctfalse}
  {\let\EndOfBibitem\relax}
\providecommand*{\mciteSetBstMidEndSepPunct}[3]{}
\providecommand*{\mciteSetBstSublistLabelBeginEnd}[3]{}
\providecommand*{\EndOfBibitem}{}
\mciteSetBstSublistMode{f}
\mciteSetBstMaxWidthForm{subitem}
{(\emph{\alph{mcitesubitemcount}})}
\mciteSetBstSublistLabelBeginEnd{\mcitemaxwidthsubitemform\space}
{\relax}{\relax}

\bibitem[Chaikin and Lubensky(1995)]{chaikin95}
P.~M. Chaikin and T.~C. Lubensky, \emph{Principles of Condensed Matter Physics}, Cambridge University Press, 1995\relax
\mciteBstWouldAddEndPuncttrue
\mciteSetBstMidEndSepPunct{\mcitedefaultmidpunct}
{\mcitedefaultendpunct}{\mcitedefaultseppunct}\relax
\EndOfBibitem
\bibitem[Kibble(1997)]{kibble97}
T.~W.~B. Kibble, \emph{Aust. J. Phys.}, 1997, \textbf{50}, 697\relax
\mciteBstWouldAddEndPuncttrue
\mciteSetBstMidEndSepPunct{\mcitedefaultmidpunct}
{\mcitedefaultendpunct}{\mcitedefaultseppunct}\relax
\EndOfBibitem
\bibitem[Pismen(1999)]{pismen99}
L.~M. Pismen, \emph{Vortices in Nonlinear Fields}, Oxford University Press, 1999\relax
\mciteBstWouldAddEndPuncttrue
\mciteSetBstMidEndSepPunct{\mcitedefaultmidpunct}
{\mcitedefaultendpunct}{\mcitedefaultseppunct}\relax
\EndOfBibitem
\bibitem[Kleman and Friedel(2008)]{kleman08}
M.~Kleman and J.~Friedel, \emph{Rev. Mod. Phys.}, 2008, \textbf{80}, 61--115\relax
\mciteBstWouldAddEndPuncttrue
\mciteSetBstMidEndSepPunct{\mcitedefaultmidpunct}
{\mcitedefaultendpunct}{\mcitedefaultseppunct}\relax
\EndOfBibitem
\bibitem[Saw \emph{et~al.}(2017)Saw, Doostmohammadi, Nier, Kocgozlu, Thampi, Toyama, Marcq, Lim, Yeomans, and Ladoux]{saw17}
T.~B. Saw, A.~Doostmohammadi, V.~Nier, L.~Kocgozlu, S.~Thampi, Y.~Toyama, P.~Marcq, C.~T. Lim, J.~M. Yeomans and B.~Ladoux, \emph{Nature}, 2017, \textbf{544}, 212--216\relax
\mciteBstWouldAddEndPuncttrue
\mciteSetBstMidEndSepPunct{\mcitedefaultmidpunct}
{\mcitedefaultendpunct}{\mcitedefaultseppunct}\relax
\EndOfBibitem
\bibitem[Maroudas-Sacks \emph{et~al.}(2021)Maroudas-Sacks, Garion, Shani-Zerbib, Livshits, Braun, and Keren]{Maroudas-Sacks21}
Y.~Maroudas-Sacks, L.~Garion, L.~Shani-Zerbib, A.~Livshits, E.~Braun and K.~Keren, \emph{Nature Physics}, 2021, \textbf{17}, 251--259\relax
\mciteBstWouldAddEndPuncttrue
\mciteSetBstMidEndSepPunct{\mcitedefaultmidpunct}
{\mcitedefaultendpunct}{\mcitedefaultseppunct}\relax
\EndOfBibitem
\bibitem[Lin \emph{et~al.}(2013)Lin, Reichhardt, Batista, and Saxena]{Shi-Zeng13}
S.-Z. Lin, C.~Reichhardt, C.~D. Batista and A.~Saxena, \emph{Phys. Rev. B}, 2013, \textbf{87}, 214419\relax
\mciteBstWouldAddEndPuncttrue
\mciteSetBstMidEndSepPunct{\mcitedefaultmidpunct}
{\mcitedefaultendpunct}{\mcitedefaultseppunct}\relax
\EndOfBibitem
\bibitem[Reichhardt and Reichhardt(2016)]{Reichhardt16}
C.~Reichhardt and C.~Reichhardt, \emph{Rep. Prog. Phys.}, 2016, \textbf{80}, 026501\relax
\mciteBstWouldAddEndPuncttrue
\mciteSetBstMidEndSepPunct{\mcitedefaultmidpunct}
{\mcitedefaultendpunct}{\mcitedefaultseppunct}\relax
\EndOfBibitem
\bibitem[Tang and Selinger(2017)]{tang17}
X.~Tang and J.~V. Selinger, \emph{Soft Matter}, 2017, \textbf{13}, 5481\relax
\mciteBstWouldAddEndPuncttrue
\mciteSetBstMidEndSepPunct{\mcitedefaultmidpunct}
{\mcitedefaultendpunct}{\mcitedefaultseppunct}\relax
\EndOfBibitem
\bibitem[Skaugen \emph{et~al.}(2018)Skaugen, Angheluta, and Vi\~nals]{audun18}
A.~Skaugen, L.~Angheluta and J.~Vi\~nals, \emph{Phys. Rev. B}, 2018, \textbf{97}, 054113\relax
\mciteBstWouldAddEndPuncttrue
\mciteSetBstMidEndSepPunct{\mcitedefaultmidpunct}
{\mcitedefaultendpunct}{\mcitedefaultseppunct}\relax
\EndOfBibitem
\bibitem[Shankar and Marchetti(2019)]{Shankar19}
S.~Shankar and M.~C. Marchetti, \emph{Phys. Rev. X}, 2019, \textbf{9}, 041047\relax
\mciteBstWouldAddEndPuncttrue
\mciteSetBstMidEndSepPunct{\mcitedefaultmidpunct}
{\mcitedefaultendpunct}{\mcitedefaultseppunct}\relax
\EndOfBibitem
\bibitem[Angheluta \emph{et~al.}(2021)Angheluta, Chen, Marchetti, and Bowick]{angheluta21}
L.~Angheluta, Z.~Chen, M.~C. Marchetti and M.~J. Bowick, \emph{New J. Phys.}, 2021, \textbf{23}, 033009\relax
\mciteBstWouldAddEndPuncttrue
\mciteSetBstMidEndSepPunct{\mcitedefaultmidpunct}
{\mcitedefaultendpunct}{\mcitedefaultseppunct}\relax
\EndOfBibitem
\bibitem[Long \emph{et~al.}(2021)Long, Tang, Selinger, and Selinger]{long21}
C.~Long, X.~Tang, R.~L. Selinger and J.~V. Selinger, \emph{Soft Matter}, 2021, \textbf{17}, 2265\relax
\mciteBstWouldAddEndPuncttrue
\mciteSetBstMidEndSepPunct{\mcitedefaultmidpunct}
{\mcitedefaultendpunct}{\mcitedefaultseppunct}\relax
\EndOfBibitem
\bibitem[Schimming and Viñals(2022)]{schimming22}
C.~D. Schimming and J.~Viñals, \emph{Soft Matter}, 2022, \textbf{18}, 2234--2244\relax
\mciteBstWouldAddEndPuncttrue
\mciteSetBstMidEndSepPunct{\mcitedefaultmidpunct}
{\mcitedefaultendpunct}{\mcitedefaultseppunct}\relax
\EndOfBibitem
\bibitem[Skogvoll \emph{et~al.}(2022)Skogvoll, Angheluta, Skaugen, Salvalaglio, and Viñals]{skogvoll22}
V.~Skogvoll, L.~Angheluta, A.~Skaugen, M.~Salvalaglio and J.~Viñals, \emph{J. Mech. Phys. Solids}, 2022, \textbf{166}, 104932\relax
\mciteBstWouldAddEndPuncttrue
\mciteSetBstMidEndSepPunct{\mcitedefaultmidpunct}
{\mcitedefaultendpunct}{\mcitedefaultseppunct}\relax
\EndOfBibitem
\bibitem[Schimming and Viñals(2023)]{schimming23}
C.~D. Schimming and J.~Viñals, \emph{Proc. R. Soc. A: Mathematical, Physical and Engineering Sciences}, 2023, \textbf{479}, 20230042\relax
\mciteBstWouldAddEndPuncttrue
\mciteSetBstMidEndSepPunct{\mcitedefaultmidpunct}
{\mcitedefaultendpunct}{\mcitedefaultseppunct}\relax
\EndOfBibitem
\bibitem[Marchetti \emph{et~al.}(2013)Marchetti, Joanny, Ramaswamy, Liverpool, Prost, Rao, and Simha]{marchetti13}
M.~C. Marchetti, J.~F. Joanny, S.~Ramaswamy, T.~B. Liverpool, J.~Prost, M.~Rao and R.~A. Simha, \emph{Rev. Mod. Phys.}, 2013, \textbf{85}, 1143\relax
\mciteBstWouldAddEndPuncttrue
\mciteSetBstMidEndSepPunct{\mcitedefaultmidpunct}
{\mcitedefaultendpunct}{\mcitedefaultseppunct}\relax
\EndOfBibitem
\bibitem[Doostmohammadi \emph{et~al.}(2018)Doostmohammadi, Ign\'{e}s-Mullol, Yeomans, and Sagu\'{e}s]{doo18}
A.~Doostmohammadi, J.~Ign\'{e}s-Mullol, J.~M. Yeomans and F.~Sagu\'{e}s, \emph{Nature Commun.}, 2018, \textbf{9}, 3246\relax
\mciteBstWouldAddEndPuncttrue
\mciteSetBstMidEndSepPunct{\mcitedefaultmidpunct}
{\mcitedefaultendpunct}{\mcitedefaultseppunct}\relax
\EndOfBibitem
\bibitem[DeCamp \emph{et~al.}(2015)DeCamp, Redner, Baskaran, Hagan, and Dogic]{DeCamp15}
S.~J. DeCamp, G.~S. Redner, A.~Baskaran, M.~F. Hagan and Z.~Dogic, \emph{Nature Mater.}, 2015, \textbf{14}, 1110--1115\relax
\mciteBstWouldAddEndPuncttrue
\mciteSetBstMidEndSepPunct{\mcitedefaultmidpunct}
{\mcitedefaultendpunct}{\mcitedefaultseppunct}\relax
\EndOfBibitem
\bibitem[Giomi(2015)]{giomi15}
L.~Giomi, \emph{Phys. Rev. X}, 2015, \textbf{5}, 031003\relax
\mciteBstWouldAddEndPuncttrue
\mciteSetBstMidEndSepPunct{\mcitedefaultmidpunct}
{\mcitedefaultendpunct}{\mcitedefaultseppunct}\relax
\EndOfBibitem
\bibitem[Doostmohammadi \emph{et~al.}(2016)Doostmohammadi, Shendruk, Thijssen, and Yeomans]{doo16b}
A.~Doostmohammadi, T.~N. Shendruk, K.~Thijssen and J.~M. Yeomans, \emph{Nature Commun.}, 2016, \textbf{8}, 15326\relax
\mciteBstWouldAddEndPuncttrue
\mciteSetBstMidEndSepPunct{\mcitedefaultmidpunct}
{\mcitedefaultendpunct}{\mcitedefaultseppunct}\relax
\EndOfBibitem
\bibitem[Lemma \emph{et~al.}(2019)Lemma, DeCamp, You, Giomi, and Dogic]{lemma19}
L.~M. Lemma, S.~J. DeCamp, Z.~You, L.~Giomi and Z.~Dogic, \emph{Soft Matter}, 2019, \textbf{15}, 3264--3272\relax
\mciteBstWouldAddEndPuncttrue
\mciteSetBstMidEndSepPunct{\mcitedefaultmidpunct}
{\mcitedefaultendpunct}{\mcitedefaultseppunct}\relax
\EndOfBibitem
\bibitem[Giomi \emph{et~al.}(2013)Giomi, Bowick, Ma, and Marchetti]{giomi13}
L.~Giomi, M.~J. Bowick, X.~Ma and M.~C. Marchetti, \emph{Phys. Rev. Lett.}, 2013, \textbf{110}, 228101\relax
\mciteBstWouldAddEndPuncttrue
\mciteSetBstMidEndSepPunct{\mcitedefaultmidpunct}
{\mcitedefaultendpunct}{\mcitedefaultseppunct}\relax
\EndOfBibitem
\bibitem[Shankar \emph{et~al.}(2018)Shankar, Ramaswamy, Marchetti, and Bowick]{shankar18}
S.~Shankar, S.~Ramaswamy, M.~C. Marchetti and M.~J. Bowick, \emph{Phys. Rev. Lett.}, 2018, \textbf{121}, 108002\relax
\mciteBstWouldAddEndPuncttrue
\mciteSetBstMidEndSepPunct{\mcitedefaultmidpunct}
{\mcitedefaultendpunct}{\mcitedefaultseppunct}\relax
\EndOfBibitem
\bibitem[Marenduzzo \emph{et~al.}(2007)Marenduzzo, Orlandini, Cates, and Yeomans]{marenduzzo07}
D.~Marenduzzo, E.~Orlandini, M.~E. Cates and J.~M. Yeomans, \emph{Phys. Rev. E}, 2007, \textbf{76}, 031921\relax
\mciteBstWouldAddEndPuncttrue
\mciteSetBstMidEndSepPunct{\mcitedefaultmidpunct}
{\mcitedefaultendpunct}{\mcitedefaultseppunct}\relax
\EndOfBibitem
\bibitem[Lazo \emph{et~al.}(2014)Lazo, Peng, Xiang, Shiyanovskii, and Lavrentovich]{lazo14}
I.~Lazo, C.~Peng, J.~Xiang, S.~V. Shiyanovskii and O.~D. Lavrentovich, \emph{Nature Commun.}, 2014, \textbf{5}, 5033\relax
\mciteBstWouldAddEndPuncttrue
\mciteSetBstMidEndSepPunct{\mcitedefaultmidpunct}
{\mcitedefaultendpunct}{\mcitedefaultseppunct}\relax
\EndOfBibitem
\bibitem[Conklin and Vi{\~n}als(2017)]{re:conklin17}
C.~Conklin and J.~Vi{\~n}als, \emph{Soft Matter}, 2017, \textbf{13}, 725--739\relax
\mciteBstWouldAddEndPuncttrue
\mciteSetBstMidEndSepPunct{\mcitedefaultmidpunct}
{\mcitedefaultendpunct}{\mcitedefaultseppunct}\relax
\EndOfBibitem
\bibitem[Genkin \emph{et~al.}(2017)Genkin, Sokolov, Lavrentovich, and Aranson]{genkin17}
M.~M. Genkin, A.~Sokolov, O.~D. Lavrentovich and I.~S. Aranson, \emph{Phys. Rev. X}, 2017, \textbf{7}, 011029\relax
\mciteBstWouldAddEndPuncttrue
\mciteSetBstMidEndSepPunct{\mcitedefaultmidpunct}
{\mcitedefaultendpunct}{\mcitedefaultseppunct}\relax
\EndOfBibitem
\bibitem[Li \emph{et~al.}(2017)Li, Armas-P\'erez, Hern\'andez-Ortiz, Arges, Liu, Mart\'inez-Gonz\'alez, Ocola, Bishop, Xie, de~Pablo, and Nealey]{li17}
X.~Li, J.~C. Armas-P\'erez, J.~P. Hern\'andez-Ortiz, C.~G. Arges, X.~Liu, J.~A. Mart\'inez-Gonz\'alez, L.~E. Ocola, C.~Bishop, H.~Xie, J.~J. de~Pablo and P.~F. Nealey, \emph{ACS Nano}, 2017, \textbf{11}, 6492--6501\relax
\mciteBstWouldAddEndPuncttrue
\mciteSetBstMidEndSepPunct{\mcitedefaultmidpunct}
{\mcitedefaultendpunct}{\mcitedefaultseppunct}\relax
\EndOfBibitem
\bibitem[Tan \emph{et~al.}(2019)Tan, Roberts, Smith, Olvera, Arteaga, Fortini, Mitchell, and Hirst]{Tan19}
A.~J. Tan, E.~Roberts, S.~A. Smith, U.~A. Olvera, J.~Arteaga, S.~Fortini, K.~A. Mitchell and L.~S. Hirst, \emph{Nature Physics}, 2019, \textbf{15}, 1033--1039\relax
\mciteBstWouldAddEndPuncttrue
\mciteSetBstMidEndSepPunct{\mcitedefaultmidpunct}
{\mcitedefaultendpunct}{\mcitedefaultseppunct}\relax
\EndOfBibitem
\bibitem[Copenhagen \emph{et~al.}(2021)Copenhagen, Alert, Wingreen, and Shaevitz]{copenhagen21}
K.~Copenhagen, R.~Alert, N.~S. Wingreen and J.~W. Shaevitz, \emph{Nature Phys.}, 2021, \textbf{17}, 211--215\relax
\mciteBstWouldAddEndPuncttrue
\mciteSetBstMidEndSepPunct{\mcitedefaultmidpunct}
{\mcitedefaultendpunct}{\mcitedefaultseppunct}\relax
\EndOfBibitem
\bibitem[Figueroa-Morales \emph{et~al.}(2022)Figueroa-Morales, Genkin, Sokolov, and Aranson]{figueroa22}
N.~Figueroa-Morales, M.~M. Genkin, A.~Sokolov and I.~S. Aranson, \emph{Commun. Phys.}, 2022, \textbf{5}, 301\relax
\mciteBstWouldAddEndPuncttrue
\mciteSetBstMidEndSepPunct{\mcitedefaultmidpunct}
{\mcitedefaultendpunct}{\mcitedefaultseppunct}\relax
\EndOfBibitem
\bibitem[Zhang \emph{et~al.}(2022)Zhang, Mozaffari, and de~Pablo]{RZhang22}
R.~Zhang, A.~Mozaffari and J.~J. de~Pablo, \emph{Sci. Adv.}, 2022, \textbf{8}, eabg9060\relax
\mciteBstWouldAddEndPuncttrue
\mciteSetBstMidEndSepPunct{\mcitedefaultmidpunct}
{\mcitedefaultendpunct}{\mcitedefaultseppunct}\relax
\EndOfBibitem
\bibitem[Serra \emph{et~al.}(2023)Serra, Lemma, Giomi, Dogic, and Mahadevan]{Serra23}
M.~Serra, L.~Lemma, L.~Giomi, Z.~Dogic and L.~Mahadevan, \emph{Nature Physics}, 2023, \textbf{19}, 1355--1361\relax
\mciteBstWouldAddEndPuncttrue
\mciteSetBstMidEndSepPunct{\mcitedefaultmidpunct}
{\mcitedefaultendpunct}{\mcitedefaultseppunct}\relax
\EndOfBibitem
\bibitem[Shankar \emph{et~al.}(2024)Shankar, Scharrer, Bowick, and Marchetti]{Shankar24}
S.~Shankar, L.~V.~D. Scharrer, M.~J. Bowick and M.~C. Marchetti, \emph{Proceedings of the National Academy of Sciences}, 2024, \textbf{121}, e2400933121\relax
\mciteBstWouldAddEndPuncttrue
\mciteSetBstMidEndSepPunct{\mcitedefaultmidpunct}
{\mcitedefaultendpunct}{\mcitedefaultseppunct}\relax
\EndOfBibitem
\bibitem[Memarian \emph{et~al.}(2024)Memarian, Hammar, Sabbir, Elias, Mitchell, and Hirst]{Memarian24}
F.~L. Memarian, D.~Hammar, M.~M.~H. Sabbir, M.~Elias, K.~A. Mitchell and L.~S. Hirst, \emph{Phys. Rev. Lett.}, 2024, \textbf{132}, 228301\relax
\mciteBstWouldAddEndPuncttrue
\mciteSetBstMidEndSepPunct{\mcitedefaultmidpunct}
{\mcitedefaultendpunct}{\mcitedefaultseppunct}\relax
\EndOfBibitem
\bibitem[Mitchell \emph{et~al.}(2024)Mitchell, Sabbir, Geumhan, Smith, Klein, and Beller]{Mitchell24}
K.~A. Mitchell, M.~M.~H. Sabbir, K.~Geumhan, S.~A. Smith, B.~Klein and D.~A. Beller, \emph{Phys. Rev. E}, 2024, \textbf{109}, 014606\relax
\mciteBstWouldAddEndPuncttrue
\mciteSetBstMidEndSepPunct{\mcitedefaultmidpunct}
{\mcitedefaultendpunct}{\mcitedefaultseppunct}\relax
\EndOfBibitem
\bibitem[Cortese \emph{et~al.}(2018)Cortese, Eggers, and Liverpool]{Cortese18}
D.~Cortese, J.~Eggers and T.~B. Liverpool, \emph{Phys. Rev. E}, 2018, \textbf{97}, 022704\relax
\mciteBstWouldAddEndPuncttrue
\mciteSetBstMidEndSepPunct{\mcitedefaultmidpunct}
{\mcitedefaultendpunct}{\mcitedefaultseppunct}\relax
\EndOfBibitem
\bibitem[Zhang \emph{et~al.}(2020)Zhang, Deserno, and Tu]{ZhangYi20}
Y.-H. Zhang, M.~Deserno and Z.-C. Tu, \emph{Phys. Rev. E}, 2020, \textbf{102}, 012607\relax
\mciteBstWouldAddEndPuncttrue
\mciteSetBstMidEndSepPunct{\mcitedefaultmidpunct}
{\mcitedefaultendpunct}{\mcitedefaultseppunct}\relax
\EndOfBibitem
\bibitem[Pismen(2013)]{Pismen13}
L.~M. Pismen, \emph{Phys. Rev. E}, 2013, \textbf{88}, 050502\relax
\mciteBstWouldAddEndPuncttrue
\mciteSetBstMidEndSepPunct{\mcitedefaultmidpunct}
{\mcitedefaultendpunct}{\mcitedefaultseppunct}\relax
\EndOfBibitem
\bibitem[Giomi \emph{et~al.}(2014)Giomi, Bowick, Mishra, Sknepnek, and Marchetti]{giomi14}
L.~Giomi, M.~J. Bowick, P.~Mishra, R.~Sknepnek and M.~C. Marchetti, \emph{Phil. Trans. R. Soc. A}, 2014, \textbf{372}, 20130365\relax
\mciteBstWouldAddEndPuncttrue
\mciteSetBstMidEndSepPunct{\mcitedefaultmidpunct}
{\mcitedefaultendpunct}{\mcitedefaultseppunct}\relax
\EndOfBibitem
\bibitem[Pismen and Sagu\'es(2017)]{Pismen17}
L.~M. Pismen and F.~Sagu\'es, \emph{Eur. Phys. J. E}, 2017, \textbf{40}, 92\relax
\mciteBstWouldAddEndPuncttrue
\mciteSetBstMidEndSepPunct{\mcitedefaultmidpunct}
{\mcitedefaultendpunct}{\mcitedefaultseppunct}\relax
\EndOfBibitem
\bibitem[R{\o}nning \emph{et~al.}(2022)R{\o}nning, Marchetti, Bowick, and Angheluta]{ronning22}
J.~R{\o}nning, C.~M. Marchetti, M.~J. Bowick and L.~Angheluta, \emph{Proc. Roy. Soc. A: Math. Phys. Eng. Sci.}, 2022, \textbf{478}, 20210879\relax
\mciteBstWouldAddEndPuncttrue
\mciteSetBstMidEndSepPunct{\mcitedefaultmidpunct}
{\mcitedefaultendpunct}{\mcitedefaultseppunct}\relax
\EndOfBibitem
\bibitem[R{\o}nning \emph{et~al.}(2023)R{\o}nning, Marchetti, and Angheluta]{Ronning23}
J.~R{\o}nning, M.~C. Marchetti and L.~Angheluta, \emph{R. Soc. Open Sci.}, 2023, \textbf{10}, 221229\relax
\mciteBstWouldAddEndPuncttrue
\mciteSetBstMidEndSepPunct{\mcitedefaultmidpunct}
{\mcitedefaultendpunct}{\mcitedefaultseppunct}\relax
\EndOfBibitem
\bibitem[Keber \emph{et~al.}(2014)Keber, Loiseau, Sanchez, DeCamp, Giomi, Bowick, Marchetti, Dogic, and Bausch]{Keber14}
F.~C. Keber, E.~Loiseau, T.~Sanchez, S.~J. DeCamp, L.~Giomi, M.~J. Bowick, M.~C. Marchetti, Z.~Dogic and A.~R. Bausch, \emph{Science}, 2014, \textbf{345}, 1135--1139\relax
\mciteBstWouldAddEndPuncttrue
\mciteSetBstMidEndSepPunct{\mcitedefaultmidpunct}
{\mcitedefaultendpunct}{\mcitedefaultseppunct}\relax
\EndOfBibitem
\bibitem[Ellis \emph{et~al.}(2018)Ellis, Pearce, Chang, Goldsztein, Giomi, and Fernandez-Nieves]{ellis18}
P.~W. Ellis, D.~J.~G. Pearce, Y.-W. Chang, G.~Goldsztein, L.~Giomi and A.~Fernandez-Nieves, \emph{Nature Physics}, 2018, \textbf{14}, 85--90\relax
\mciteBstWouldAddEndPuncttrue
\mciteSetBstMidEndSepPunct{\mcitedefaultmidpunct}
{\mcitedefaultendpunct}{\mcitedefaultseppunct}\relax
\EndOfBibitem
\bibitem[Pearce \emph{et~al.}(2021)Pearce, Nambisan, Ellis, Fernandez-Nieves, and Giomi]{Pearce21b}
D.~J.~G. Pearce, J.~Nambisan, P.~W. Ellis, A.~Fernandez-Nieves and L.~Giomi, \emph{Phys. Rev. Lett.}, 2021, \textbf{127}, 197801\relax
\mciteBstWouldAddEndPuncttrue
\mciteSetBstMidEndSepPunct{\mcitedefaultmidpunct}
{\mcitedefaultendpunct}{\mcitedefaultseppunct}\relax
\EndOfBibitem
\bibitem[Head \emph{et~al.}(2024)Head, Dor\'{e}, Keogh, Bonn, Negro, Marenduzzo, Doostmohommadi, Thijssen, L'{o}pez-Le'{o}n, and Shendruk]{Head24}
L.~C. Head, C.~Dor\'{e}, R.~R. Keogh, L.~Bonn, G.~Negro, D.~Marenduzzo, A.~Doostmohommadi, K.~Thijssen, T.~L'{o}pez-Le'{o}n and T.~N. Shendruk, \emph{Nature Physics}, 2024, \textbf{20}, 492--500\relax
\mciteBstWouldAddEndPuncttrue
\mciteSetBstMidEndSepPunct{\mcitedefaultmidpunct}
{\mcitedefaultendpunct}{\mcitedefaultseppunct}\relax
\EndOfBibitem
\bibitem[Vafa(2022)]{Vafa22}
F.~Vafa, \emph{Soft Matter}, 2022, \textbf{18}, 8087--8097\relax
\mciteBstWouldAddEndPuncttrue
\mciteSetBstMidEndSepPunct{\mcitedefaultmidpunct}
{\mcitedefaultendpunct}{\mcitedefaultseppunct}\relax
\EndOfBibitem
\bibitem[de~Gennes(1975)]{deGennes75}
P.~G. de~Gennes, \emph{The Physics of Liquid Crystals}, Oxford University Press, 1975\relax
\mciteBstWouldAddEndPuncttrue
\mciteSetBstMidEndSepPunct{\mcitedefaultmidpunct}
{\mcitedefaultendpunct}{\mcitedefaultseppunct}\relax
\EndOfBibitem
\bibitem[Halperin(1981)]{halperin81}
B.~I. Halperin, \emph{Physics of Defects}, North-Holland Pub. Co., 1981\relax
\mciteBstWouldAddEndPuncttrue
\mciteSetBstMidEndSepPunct{\mcitedefaultmidpunct}
{\mcitedefaultendpunct}{\mcitedefaultseppunct}\relax
\EndOfBibitem
\bibitem[Liu and Mazenko(1992)]{liu92}
F.~Liu and G.~F. Mazenko, \emph{Phys. Rev. B}, 1992, \textbf{46}, 5963\relax
\mciteBstWouldAddEndPuncttrue
\mciteSetBstMidEndSepPunct{\mcitedefaultmidpunct}
{\mcitedefaultendpunct}{\mcitedefaultseppunct}\relax
\EndOfBibitem
\bibitem[Mazenko and Wickham(1997)]{mazenko97}
G.~F. Mazenko and R.~A. Wickham, \emph{Phys. Rev. E}, 1997, \textbf{57}, 2539\relax
\mciteBstWouldAddEndPuncttrue
\mciteSetBstMidEndSepPunct{\mcitedefaultmidpunct}
{\mcitedefaultendpunct}{\mcitedefaultseppunct}\relax
\EndOfBibitem
\bibitem[Mazenko(1999)]{mazenko99}
G.~F. Mazenko, \emph{Phys. Rev. E}, 1999, \textbf{59}, 1574\relax
\mciteBstWouldAddEndPuncttrue
\mciteSetBstMidEndSepPunct{\mcitedefaultmidpunct}
{\mcitedefaultendpunct}{\mcitedefaultseppunct}\relax
\EndOfBibitem
\bibitem[Beris and Edwards(1994)]{beris94}
A.~N. Beris and B.~J. Edwards, \emph{Thermodynamics of flowing systems}, Oxford University Press, 1994\relax
\mciteBstWouldAddEndPuncttrue
\mciteSetBstMidEndSepPunct{\mcitedefaultmidpunct}
{\mcitedefaultendpunct}{\mcitedefaultseppunct}\relax
\EndOfBibitem
\bibitem[Leslie(1966)]{leslie66}
F.~M. Leslie, \emph{J. Mech. Appl. Math.}, 1966, \textbf{19}, 357\relax
\mciteBstWouldAddEndPuncttrue
\mciteSetBstMidEndSepPunct{\mcitedefaultmidpunct}
{\mcitedefaultendpunct}{\mcitedefaultseppunct}\relax
\EndOfBibitem
\bibitem[T\'oth \emph{et~al.}(2002)T\'oth, Denniston, and Yeomans]{toth02}
G.~T\'oth, C.~Denniston and J.~M. Yeomans, \emph{Phys. Rev. Lett.}, 2002, \textbf{88}, 105504\relax
\mciteBstWouldAddEndPuncttrue
\mciteSetBstMidEndSepPunct{\mcitedefaultmidpunct}
{\mcitedefaultendpunct}{\mcitedefaultseppunct}\relax
\EndOfBibitem
\bibitem[Vromans and Giomi(2016)]{vromans16}
A.~J. Vromans and L.~Giomi, \emph{Soft Matter}, 2016, \textbf{12}, 6490\relax
\mciteBstWouldAddEndPuncttrue
\mciteSetBstMidEndSepPunct{\mcitedefaultmidpunct}
{\mcitedefaultendpunct}{\mcitedefaultseppunct}\relax
\EndOfBibitem
\bibitem[Norton \emph{et~al.}(2018)Norton, Baskaran, Opathalage, Langeslay, Fraden, Baskaran, and Hagan]{norton18}
M.~M. Norton, A.~Baskaran, A.~Opathalage, B.~Langeslay, S.~Fraden, A.~Baskaran and M.~F. Hagan, \emph{Phys. Rev. E}, 2018, \textbf{97}, 012702\relax
\mciteBstWouldAddEndPuncttrue
\mciteSetBstMidEndSepPunct{\mcitedefaultmidpunct}
{\mcitedefaultendpunct}{\mcitedefaultseppunct}\relax
\EndOfBibitem
\bibitem[Schimming \emph{et~al.}(2023)Schimming, Reichhardt, and Reichhardt]{schimming23b}
C.~D. Schimming, C.~J.~O. Reichhardt and C.~Reichhardt, \emph{Phys. Rev. E}, 2023, \textbf{108}, L012602\relax
\mciteBstWouldAddEndPuncttrue
\mciteSetBstMidEndSepPunct{\mcitedefaultmidpunct}
{\mcitedefaultendpunct}{\mcitedefaultseppunct}\relax
\EndOfBibitem
\bibitem[Schimming \emph{et~al.}(2024)Schimming, Reichhardt, and Reichhardt]{schimming24}
C.~D. Schimming, C.~J.~O. Reichhardt and C.~Reichhardt, \emph{Phys. Rev. Lett.}, 2024, \textbf{132}, 018301\relax
\mciteBstWouldAddEndPuncttrue
\mciteSetBstMidEndSepPunct{\mcitedefaultmidpunct}
{\mcitedefaultendpunct}{\mcitedefaultseppunct}\relax
\EndOfBibitem
\bibitem[Opathalage \emph{et~al.}(2019)Opathalage, Norton, Juniper, Langeslay, Aghvami, Fraden, and Dogic]{opathalage19}
A.~Opathalage, M.~M. Norton, M.~P.~N. Juniper, B.~Langeslay, S.~A. Aghvami, S.~Fraden and Z.~Dogic, \emph{Proc. Natl. Acad. Sci. (USA)}, 2019, \textbf{116}, 4788--4797\relax
\mciteBstWouldAddEndPuncttrue
\mciteSetBstMidEndSepPunct{\mcitedefaultmidpunct}
{\mcitedefaultendpunct}{\mcitedefaultseppunct}\relax
\EndOfBibitem
\bibitem[Sokolov \emph{et~al.}(2024)Sokolov, Katuri, de~Pablo, and Snezhko]{sokolov24}
A.~Sokolov, J.~Katuri, J.~J. de~Pablo and A.~Snezhko, \emph{Synthetic active liquid crystals powered by acoustic waves}, 2024, \url{https://arxiv.org/abs/2403.17268}\relax
\mciteBstWouldAddEndPuncttrue
\mciteSetBstMidEndSepPunct{\mcitedefaultmidpunct}
{\mcitedefaultendpunct}{\mcitedefaultseppunct}\relax
\EndOfBibitem
\bibitem[Hemingway \emph{et~al.}(2016)Hemingway, Mishra, Marchetti, and Fielding]{Hemingway16}
E.~J. Hemingway, P.~Mishra, M.~C. Marchetti and S.~M. Fielding, \emph{Soft Matter}, 2016, \textbf{12}, 7943--7952\relax
\mciteBstWouldAddEndPuncttrue
\mciteSetBstMidEndSepPunct{\mcitedefaultmidpunct}
{\mcitedefaultendpunct}{\mcitedefaultseppunct}\relax
\EndOfBibitem
\bibitem[Repula \emph{et~al.}(2024)Repula, Gates, Cameron, and Smalyukh]{repula24}
A.~Repula, C.~Gates, J.~C. Cameron and I.~I. Smalyukh, \emph{Commun. Materials}, 2024, \textbf{5}, 37\relax
\mciteBstWouldAddEndPuncttrue
\mciteSetBstMidEndSepPunct{\mcitedefaultmidpunct}
{\mcitedefaultendpunct}{\mcitedefaultseppunct}\relax
\EndOfBibitem
\bibitem[Pearce and Kruse(2021)]{pearce21}
D.~J.~G. Pearce and K.~Kruse, \emph{Soft Matter}, 2021, \textbf{17}, 7408\relax
\mciteBstWouldAddEndPuncttrue
\mciteSetBstMidEndSepPunct{\mcitedefaultmidpunct}
{\mcitedefaultendpunct}{\mcitedefaultseppunct}\relax
\EndOfBibitem
\bibitem[Zushi \emph{et~al.}(2024)Zushi, Schimming, and Takeuchi]{Zushi24}
Y.~Zushi, C.~D. Schimming and K.~A. Takeuchi, \emph{Phys. Rev. Res.}, 2024, \textbf{6}, 023284\relax
\mciteBstWouldAddEndPuncttrue
\mciteSetBstMidEndSepPunct{\mcitedefaultmidpunct}
{\mcitedefaultendpunct}{\mcitedefaultseppunct}\relax
\EndOfBibitem
\bibitem[\ifmmode~\check{C}\else \v{C}\fi{}opar \emph{et~al.}(2019)\ifmmode~\check{C}\else \v{C}\fi{}opar, Aplinc, Kos, \ifmmode~\check{Z}\else \v{Z}\fi{}umer, and Ravnik]{Copar19}
S.~\ifmmode~\check{C}\else \v{C}\fi{}opar, J.~Aplinc, i.~c.~v. Kos, S.~\ifmmode~\check{Z}\else \v{Z}\fi{}umer and M.~Ravnik, \emph{Phys. Rev. X}, 2019, \textbf{9}, 031051\relax
\mciteBstWouldAddEndPuncttrue
\mciteSetBstMidEndSepPunct{\mcitedefaultmidpunct}
{\mcitedefaultendpunct}{\mcitedefaultseppunct}\relax
\EndOfBibitem
\bibitem[Kralj \emph{et~al.}(2023)Kralj, Ravnik, and Kos]{kralj23}
N.~Kralj, M.~Ravnik and i.~c.~v. Kos, \emph{Phys. Rev. Lett.}, 2023, \textbf{130}, 128101\relax
\mciteBstWouldAddEndPuncttrue
\mciteSetBstMidEndSepPunct{\mcitedefaultmidpunct}
{\mcitedefaultendpunct}{\mcitedefaultseppunct}\relax
\EndOfBibitem
\bibitem[Duclos \emph{et~al.}(2020)Duclos, Adkins, Banerjee, Peterson, Varghese, Kolvin, Baskaran, Pelcovits, Powers, Baskaran, Toschi, Hagan, Streichan, Vitelli, Beller, and Dogic]{duclos20}
G.~Duclos, R.~Adkins, D.~Banerjee, M.~S.~E. Peterson, M.~Varghese, I.~Kolvin, A.~Baskaran, R.~A. Pelcovits, T.~R. Powers, A.~Baskaran, F.~Toschi, M.~F. Hagan, S.~J. Streichan, V.~Vitelli, D.~A. Beller and Z.~Dogic, \emph{Science}, 2020, \textbf{367}, 112--1124\relax
\mciteBstWouldAddEndPuncttrue
\mciteSetBstMidEndSepPunct{\mcitedefaultmidpunct}
{\mcitedefaultendpunct}{\mcitedefaultseppunct}\relax
\EndOfBibitem
\bibitem[Binysh \emph{et~al.}(2020)Binysh, Kos, \u{C}opar, Ravnik, and Alexander]{binysh20}
J.~Binysh, {\u{Z}}.~Kos, S.~\u{C}opar, M.~Ravnik and G.~P. Alexander, \emph{Phys. Rev. Lett.}, 2020, \textbf{124}, 088001\relax
\mciteBstWouldAddEndPuncttrue
\mciteSetBstMidEndSepPunct{\mcitedefaultmidpunct}
{\mcitedefaultendpunct}{\mcitedefaultseppunct}\relax
\EndOfBibitem
\bibitem[Houston and Alexander(2022)]{Houston22}
A.~J.~H. Houston and G.~P. Alexander, \emph{Phys. Rev. E}, 2022, \textbf{105}, L062601\relax
\mciteBstWouldAddEndPuncttrue
\mciteSetBstMidEndSepPunct{\mcitedefaultmidpunct}
{\mcitedefaultendpunct}{\mcitedefaultseppunct}\relax
\EndOfBibitem
\bibitem[Walker(2018)]{walker18}
S.~W. Walker, \emph{SIAM J. Sci. Comput.}, 2018, \textbf{40}, C234--C257\relax
\mciteBstWouldAddEndPuncttrue
\mciteSetBstMidEndSepPunct{\mcitedefaultmidpunct}
{\mcitedefaultendpunct}{\mcitedefaultseppunct}\relax
\EndOfBibitem
\bibitem[Blow \emph{et~al.}(2014)Blow, Thampi, and Yeomans]{blow14}
M.~L. Blow, S.~P. Thampi and J.~M. Yeomans, \emph{Phys. Rev. Lett.}, 2014, \textbf{113}, 248303\relax
\mciteBstWouldAddEndPuncttrue
\mciteSetBstMidEndSepPunct{\mcitedefaultmidpunct}
{\mcitedefaultendpunct}{\mcitedefaultseppunct}\relax
\EndOfBibitem
\end{mcitethebibliography}
\bibliographystyle{rsc} 

\end{document}